\documentclass[10pt,twocolumn,aps,prx,superscriptaddress,longbibliography,eprint]{revtex4-2}

\usepackage{graphicx}
\usepackage{amsmath}
\usepackage{amssymb}
\usepackage{float}
\usepackage{braket}   
\usepackage{enumitem}
\usepackage[dvipsnames]{xcolor}
\usepackage[colorlinks=true, urlcolor=blue, linkcolor=blue,citecolor=blue, hypertexnames=false]{hyperref}
\usepackage[detect-weight=true,separate-uncertainty=true]{siunitx}
\usepackage{comment}
\usepackage[normalem]{ulem}

\usepackage{cleveref}
\crefname{equation}{Eq.}{Eqs.}
\Crefname{equation}{Equation}{Equations}
\crefname{figure}{Fig.}{Figs.}
\Crefname{figure}{Figure}{Figures}
\crefname{section}{Sec.}{Sects.}
\Crefname{section}{Section}{Sections}
\crefname{table}{Table}{Tables}
\crefname{appendix}{Appendix}{Apps.}
\Crefname{appendix}{Appendix}{Apps.}

\newcommand{\ha}{\hat{a}}
\newcommand{\had}{\hat{a}^\dagger}
\newcommand{\hb}{\hat{b}}

\newcommand{\hn}{\hat{n}}

\newcommand{\hH}{\hat{H}}

\newcommand{\circled}[1]{\textcircled{\raisebox{0ex}{\scriptsize #1}}}

\begin{document}

\title{Measurement-induced state transitions in multi-qubit transmon processors}

\author{Baptiste Hoyau}
\author{Alexander McDonald}
\author{Boris M. Varbanov}
\author{Manuel H. Mu\~noz-Arias}
\altaffiliation{Peresent address: Quantum Algorithms and Applications Collaboratory, Sandia National Laboratories, Livermore, CA 94550, USA}
\affiliation{Institut Quantique and D\'epartement de Physique, Universit\'e de Sherbrooke, Sherbrooke J1K 2R1 QC, Canada}
\author{Alexandre Blais}
\affiliation{Institut Quantique and D\'epartement de Physique, Universit\'e de Sherbrooke, Sherbrooke J1K 2R1 QC, Canada}

\date{\today}

\begin{abstract}
Dispersive readout of the transmon qubit in circuit QED is known to lose its quantum non-demolition character at small to moderate measurement drive amplitudes.
This phenomenon is understood to originate from  Laundau-Zener transitions at accidental multi-photon resonances, where $n$ drive photons can promote the transmon by $m$ levels. This interpretation has been shown to be in agreement with experiments characterizing the dispersive readout of a single transmon. The impact of these measurement-induced state transition (MIST) of a transmon embedded in a multi-qubit chip, however, remains largely unexplored. Here, we show that the presence of other components, such as qubits and couplers, can affect the MIST threshold of a measured transmon. To arrive at these results, we present a general method to characterize measurement-induced transition when the qubit under readout is coupled to other circuit elements, a ubiquitous situation in circuit QED-based quantum processors. As an example, we consider the case of two transmon qubits, and we show that the spectator qubit can be impacted by the measurement-induced transition of the readout qubit and, conversely, that the presence of the spectator qubit can lower the MIST threshold of the readout qubit. Finally, we explore how adding a coupler mode between the two qubits further modifies these effects.  
\end{abstract}

\maketitle

\section{Introduction}
Circuit quantum electrodynamics (cQED) provides a versatile platform for fault-tolerant quantum computation~\cite{Blais2023rmp}. The ability to engineer various types of qubits~\cite{Koch2007transmon, Schreier2008Transmon, Manucharyan2009-Fluxonium, Paik2011Transmon3D, Barends2013Xmon, Brooks20130Pi, Nguyen2019Fluxonium, Gyenis20210pi} and to tailor the hardware to different quantum error correcting codes~\cite{Leghtas2013Cat, Grimm2020Cat, CampagneIbarcq2020GKP, Kubica2023ErasureQubits, Sivak2023GKP, Chou2024-DualRail} has made this architectures particularly flexible. Furthermore, there has been significant progress in the fidelity and duration of single-qubit and two-qubit gates~\cite{Barends2014Gates, Rol16SQGate, Barends2019Diabatic, Rol2019Netzero, Hong2019Parametric, Foxen2020Continious, Negirneac2021FastNetZero, Li2023Gates, Jurcevic2021Gates, Stehlik2021Gates, Sung2021Gates, Wei2022Gates}. In contrast, despite notable progress in qubit readout \cite{Swiadek2024readout,Spring2025,Sunada2022readout,Heinsoo2018readout}, further improvements in readout fidelity and speed remain a central challenge for superconducting circuits. Characterizing the constrains on the fidelity, duration and quantum non-demolition (QND) nature of dispersive readout is therefore crucial for realizing a large-scale fault-tolerant quantum computer using superconducting-qubit processors.

To this end, there have been intense efforts to elucidate the experimentally observed breakdown of the dispersive readout of a single transmon qubit~\cite{Boissonneault2008,Boissonneault2009,Verney2019,Lescanne2019,Petrescu2020,Hanai2021,  Thorbeck2024,Sank2016mist, Walter2017readout, Shillito2022, Cohen2023, Khezri2023mist, Dumas2024, Wang2025exp, Fechant2025exp, Xia2025-tj, Dai2025DriveInduced, Connolly2025multimode_exp}. It is now understood that multi-photon transmon-resonator resonances are at the origin of this behavior, something which has been referred to as drive-induced unwanted state transitions (DUST), measurement-induced state transitions (MIST) and ionization in the literature.
These photon-number-dependent resonances can be characterized using tools such as branch analysis~\cite{Boissonneault2010_ImproveNonlinear,Shillito2022} and Floquet branch analysis~\cite{Cohen2023,Dumas2024, xiao2024diagrammaticmethodcomputeeffective}, leading to remarkable agreement between theory and experiments \cite{Dumas2024,Fechant2025exp,Wang2025exp,Xia2025-tj,Dai2025DriveInduced,Connolly2025multimode_exp}. The insight provided by these tools has also lead to the design of novel circuits and readout schemes that are more robust to unwanted transitions~\cite{Chapple2025Balanced,Chapple2025Robustness,Mori2025,Beaulieu2026}. Moreover, the Floquet branch analysis makes it clear that these resonances are not specific to readout of a transmon qubit, but rather are a generic feature of any anharmonic system subject to a strong  drive~\cite{Breuer1989,Xia2025-tj, Nesterov2024MIST, Chapple2026FluxoniumMIST}.

\begin{figure}[t]
    \centering\includegraphics[width=0.8\linewidth]{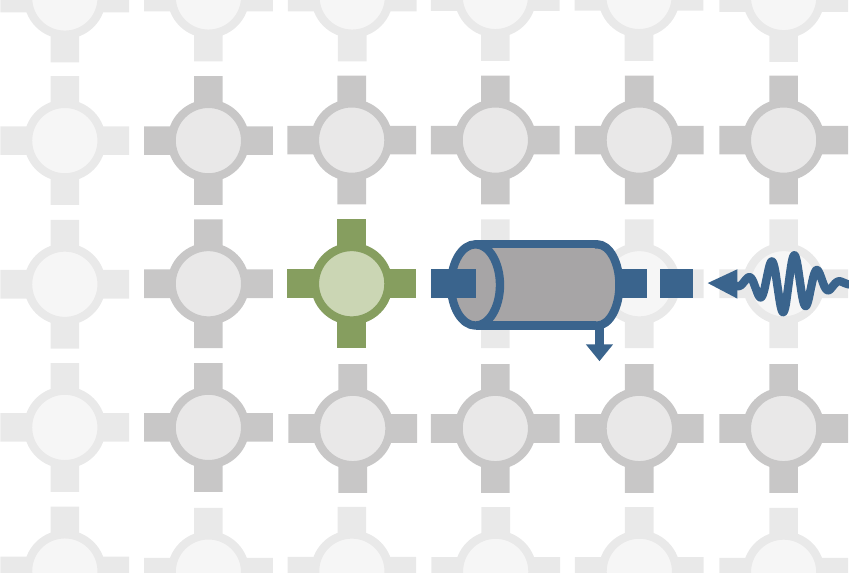}
    \caption{Schematic of a QPU focusing on a qubit (green) capacitively that is coupled to a driven readout resonator (blue) and to several spectator qubits (gray) that are themselves coupled to each other.
    }
    \label{fig:Full_system}
\end{figure}

These advances focus primarily on characterizing the readout of a single qubit. Quantum processor units (QPUs), however, consist of many coupled qubits, as depicted schematically in \cref{fig:Full_system}. Optimizing the readout of a single, well-isolated qubit does not necessarily guarantee optimal performance when that qubit is embedded in a QPU; see Ref.~\cite{Wang2025exp} for an example where the presence of a neighboring qubit modified the onset of unwanted transitions during readout. Given the important role that MIST plays in limiting the readout fidelity and duration of superconducting qubits, a framework is needed to understand the impact of strong drives and strong non-linearities in these multi-mode systems. In this work, we introduce such a tool, which analyses the impact of spectators on the measurement of a strongly-driven target qubit. The methodology generalizes the branch analysis and Floquet branch analysis methods~\cite{Dumas2024} to multiple modes. It identifies and allows one to track spectator-enabled multi-photon resonances, while also being insensitive to the trivial weak dressing of the qubit by these modes.

This paper is organized as follows. In~\cref{sec:nutshell}, we give a brief review of the simpler case of the measurement induced transition of a single qubit using the Floquet branch analysis. Next, in~\cref{sec:Multimode_branch_analysis}, we introduce a method to explore how the presence of spectator modes can induce MIST in the measured qubit mode. This method is general, free of assumptions on the nature of the spectator modes (e.g., transmon, coupler, fluxonium, resonator or a two-level system) and on the type of coupling between the modes. In~\cref{sec:single_spectator} we present the results obtained when applying this method on the simplest example: a transmon qubit capacitively coupled to a single spectator transmon. In~\cref{sec:STTC} we use our method to consider a situation where a pair of qubits are interacting via a single-transmon tunable coupler, to explore the impact that couplers might have on the measurement. Finally, we summarize our findings in~\cref{sec:conclusion}.

\section{MIST in transmon in a nutshell} \label{sec:nutshell}

In this section, we briefly review the basic mechanism leading to transmon measurement-induced state transition, focusing on a single transmon in interaction with a resonator mode \cite{Breuer1989, Sank2016mist, Shillito2022, Cohen2023, Khezri2023mist, Dumas2024}. We focus on the semiclassical Floquet branch analysis method of Ref.~\cite{Dumas2024}, which has been successfully used in understanding experimental data \cite{Dumas2024,Wang2025exp,Fechant2025exp, Xia2025-tj}. For a more complete discussion of this method, we refer the reader to Ref.~\cite{Dumas2024}.

The starting point of the semiclassical analysis is the fully-quantum Hamiltonian describing the transmon, the resonator, and their interaction ($\hbar = 1$)
\begin{equation}\label{eq:H quantum transmon resonator}
    \hH_{qr} = \omega_r\had\ha + \hH_q - ig\hn_q(\ha-\had) + \hH_d(t),
\end{equation}
where $\omega_r$ is the bare resonator frequency, $g$ the transmon-resonator coupling strength, and $\ha$ the annihilation operator of the resonator. In addition,
$\hH_q = 4E_C\hn_q^2 - E_J\cos\hat{\varphi}_q$ is the transmon Hamiltonian with $E_C$ the charging energy, $E_J$ the Josephson energy, and where we have taken the gate charge $n_g = 0$. In previous studies focusing on a single transmon, it was shown that fluctuations in the gate charge have an impact on MIST \cite{Cohen2023,Khezri2023mist,Dumas2024,Fechant2025exp}. However, because variations in gate charge do not modify the qualitative discussion in an essential
way when moving from a single qubit to multiple qubits, we take $n_g = 0$ below;
see \cref{appendix:impact of gate charge} for a brief discussion. The last term of \cref{eq:H quantum transmon resonator} is the drive on the resonator which we take to be
\begin{equation}\label{eq:res_drive}
    \hH_d(t) = i\epsilon_d \sin(\omega_dt) (\had-\ha),
\end{equation}
with $\epsilon_d$ the drive amplitude and $\omega_d$ the drive frequency chosen to be close to the cavity frequency. 

In the dispersive limit, during readout the resonator is approximately in a coherent state. Making a displacement transformation on the resonator $\ha \to \ha + \alpha(t)$, with $\alpha(t)$ chosen such as to cancel the drive, leads to an effective transmon drive $\hH(t) = \hH_q +\mathcal{E}_q(t)\cos(\omega_dt)\hn_q$ of amplitude
\begin{equation}
    \mathcal{E}_q(t) = \frac{2g\epsilon_d}{\kappa}(1-e^{-\kappa t/2}),
\end{equation}
with $\kappa$ the cavity decay rate \cite{Dumas2024}.
The resonator in this displaced frame remains in its vacuum state and, neglecting quantum fluctuations, can be ignored altogether. Crucially, $\mathcal{E}_q(t)$ varies much more slowly than the drive frequency $\omega_d$, and can be considered constant over the period of the drive $T = 2\pi/\omega_d$.  Thus, for any fixed $\mathcal{E}_q$, the resulting transmon Hamiltonian
\begin{align}
    \hH_{\mathcal{E}_q}(t) = \hat{H}_q  + \mathcal{E}_q \cos (\omega_d t) \hn_q
\end{align}
is periodic $\hH_{\mathcal{E}q}(t+T) = \hat{H}_{\mathcal{E}_q}(t)$, and can thus be analyzed using the tools of Floquet theory \cite{Dumas2024,Breuer1989}.

The drive-amplitude-dependent Floquet modes $|i_q[\mathcal{E}_q]\rangle$ are eigenvectors of the one-period propagator with eigenvalues $e^{-i\epsilon_{i_q}[\mathcal{E}_q]T}$ where 
$\epsilon_{i_q}[\mathcal{E}_q]$ are the Floquet quasi-energies \cite{Grifoni1998}. At zero drive, the Floquet modes coincide with the transmon eigenstates $|i_q[\mathcal{E}_q = 0]\rangle \equiv |i_q\rangle$. From this starting point, the Floquet modes at non-zero $\mathcal{E}_q$ are labeled by adiabatically increasing the amplitude of the drive 
and hence the effective photon number. The collection of Floquet states $B_{i_q} \equiv  \{|i_q[\mathcal{E}_q]\rangle \: | \: \forall \: \mathcal{E}_q \} $ obtained from the bare transmon state $\ket{i_q}$ as its starting point is referred to as a branch. Each branch represents the transmon states dressed by adiabatically adding the drive photons populating the resonator. 

This dressing comes in two flavors. Recall that we are interested in a dispersive measurement where the drive is off-resonant from the transmon frequency. At low drive amplitude, perturbation theory implies that we should expect a ``trivial" dressing of the Floquet states where a Floquet mode remains close to its corresponding bare state $|i_q \rangle$:  $|i_q[\mathcal{E}_q]\rangle \approx |i_q \rangle$ with the approximate equality sign slowly becoming less valid as the drive amplitude is increased. In the average transmon population of each branch,
\begin{equation}
    \langle\hat{N}_q\rangle_{i_q} = \sum_{j_q}j_q|\langle j_q|i_q[\mathcal{E}_q]\rangle|^2,
    \label{eq:transmon_population}
\end{equation}
this dressing manifests itself in a slow change
in $\langle\hat{N}_q\rangle_{i_q}$ versus $\mathcal{E}_q$. This trivial dressing at low drive strength is a consequence of the transverse nature of the qubit-resonator coupling in \cref{eq:H quantum transmon resonator} and can be captured by second-order perturbation theory; see \cref{fig:qubit} in \cref{appendix:single mode ionization} for an example of average transmon population vs drive amplitude.

The other type of dressing is due to the presence of multi-photon resonances in the ac-Stark shifted qubit-resonator spectrum \cite{Breuer1989, Sank2016mist, Shillito2022, Cohen2023, Khezri2023mist, Dumas2024}. When such a resonance occur at a given drive strength $\mathcal{E}_q$, the
perturbative description breaks down: the drive can resonantly cause transitions between different qubit branches. In contrast to the trivial dressing discussed above, this leads to avoided crossings in the quasienergy spectrum and to dramatic changes in the branch populations with a behavior that depends on the sign of the qubit-resonator detuning \cite{Cohen2023,Dumas2024}. For instance, for $\omega_r > \omega_q$, where $\omega_q$ is the qubit frequency, the branches involved in the multi-photon resonance ``swap" their population at the location of the avoided crossing in the quasi-energy spectrum.
These swappings have a dynamical consequence. Starting in a qubit eigenstate $|i_q\rangle$ and adiabatically increasing the drive strength $\mathcal{E}_q$ to the threshold where a swap occurs, the qubit's population  experiences a sudden jump to another state, i.e., there is a measurement-induced state transition. As a result, the Floquet branch analysis serves as a tool to quantitatively pinpoint at which drive strength readout is expected to breakdown due to the presence of multi-photon resonances.

\section{Multimode branch analysis}
\label{sec:Multimode_branch_analysis}

Extending the branch analysis to account for spectators is not straightforward. We aim to simultaneously capture if and when their presence leads to drastic changes in the qubit state, while ignoring any trivial dressing arising from coupling multiple modes. As we now explain, doing so requires comparing two labeling methods, which can be interpreted as running two different gedanken experiments.

The generic situation we consider is sketched in Fig. \ref{fig:Full_system}. It consists of a qubit of Hamiltonian $\hH_q$ coupled to one or more spectator modes of Hamiltonian $\hH_s$. Accounting for the coupling of the qubit to a readout resonator, the full Hamiltonian of this  system is given by
\begin{equation}
    \hH = \omega_r\had\ha + \hH_{qs} - ig\hn_q(\ha-\had) + \hH_d(t),
    \label{eq:H_qubit+spec_general}
\end{equation}
where $\hH_d(t)$ is defined in \cref{eq:res_drive} and $\hH_{qs}$ is the Hamiltonian describing the qubit mode, the spectator modes, and their respective couplings 
\begin{equation}
    \hH_{qs} =  \sum_{i=q,s_0,\dots}\hH_{i} + \sum_{i<j}J_{ij}\hn_i\hn_j.
\end{equation}
Note that throughout the article, we use two different convention for the coupling, $J_{ij}$ and $g_{ij} = J_{ij}/n_{zpf,i}n_{zpf,j}$ where $n_{zpf,i}$ is the charge zero point fluctuation of the mode $i$. Further, in writing ~\cref{eq:H_qubit+spec_general} we assume a charge-charge coupling between the modes. However, the labeling method applies to other couplings, i.e.,~phase-phase coupling.

As in the Floquet branch analysis in the single-qubit case, the labeling scheme is based on the semi-classical approximation where the effects of the resonator are entirely captured by a direct drive on the qubit. It will be useful to vary the qubit-spectator coupling between zero and the target value $J_{ij}$. We thus introduce a scaling parameter $\lambda$ such that $J_{ij} \to \lambda J_{ij}$ with $\lambda \in [0,1]$. The Hamiltonian of interest in the semiclassical approximation then takes the form
\begin{equation}
    \hat{H}_{ \lambda , \mathcal{E}_q }(t)
    =
    \sum_{i=q,s_0,\dots}\hat{H}_i 
    + \lambda\sum_{i<j}J_{ij}\hn_i\hn_j
    + \mathcal{E}_q \cos(\omega_d t) \hn_q.
    \label{eq:H_multimode}
\end{equation}
As discussed in \cref{sec:nutshell}, the drive amplitude $\mathcal{E}_q$ is related to the number of photons in the resonator. For this reason, below we use the terms drive amplitude and photon number interchangeably.

\begin{figure}[t]
    \centering
    \includegraphics[width=\linewidth]{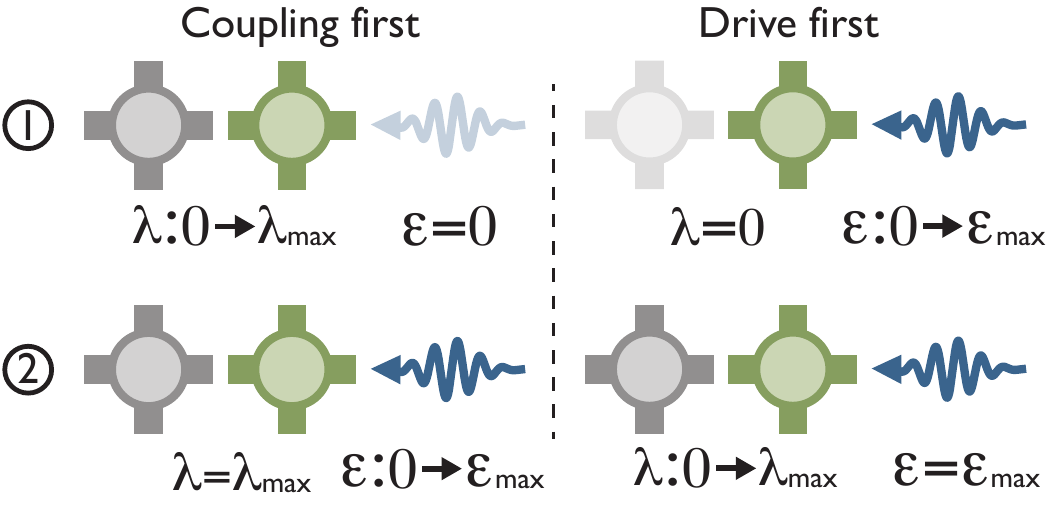}
    \caption{Schematic of the two Floquet branch analysis methods used to study the impact of a spectator mode (gray) on the MIST of a qubit (green) during dispersive readout: the coupling-first and the drive-first approaches. These two approaches proceed in two steps (labeled \circled{1} and \circled{2}).
    As their names suggest, in the coupling-first (drive-first) approach the coupling $\lambda$ (drive amplitude $\epsilon$) is first adiabatically increased from 0 to its target value while the other parameter is held at zero. In a second step, that remaining parameter is then ramped up.
    }
    \label{fig:fig2}
\end{figure}

Our objective is to understand (i) how the Floquet modes and quasienergies of the Hamiltonian with the spectators, $\hat{H}_{\lambda = 1 , \mathcal{E}_q}(t)$, differ from that of the Hamiltonian without any coupling, $\hat{H}_{ \lambda = 0, \mathcal{E}_q}(t)$, and (ii) at what drive strength do drastic differences between the Floquet modes emerge due to the presence of the spectators. To reach these goals we label the Floquet branches using two different approaches: coupling-first and drive-first Floquet branch analysis, see \cref{fig:fig2}.

\subsection{Coupling-first Floquet branch analysis}

We begin with the uncoupled $\lambda = 0$ and undriven $\mathcal{E}_q = 0$ eigenvectors $|i_q,j_s\rangle$ of the qubit-spectator system. Here $j_s$ is a shorthand for the indices of the spectators, e.g., $|j_s\rangle = |j_{s_0},j_{s_1},\dots\rangle$. We then adiabatically increases the coupling $\lambda = 0 \to \lambda = 1$ and use the Floquet branch analysis to obtain the state $|i_q, j_s\rangle \to|\overline{i_q, j_s}\rangle$ with the corresponding dressed energy $E_{\overline{i_q,j_s}}$. Upon reintroducing the drive on the qubit, this allows us to write Eq. (\ref{eq:H_multimode}) as 
\begin{align}
    \hat{H}_{  \lambda = 1, \mathcal{E}_q }(t)
    =
    \sum_{i_q, j_s}
    E_{\overline{i_q,j_s}} |\overline{i_q, j_s}\rangle \langle \overline{i_q, j_s} |
    + \mathcal{E}_q \cos(\omega_d t) \hn_q.
\label{eq:adiabatic}
\end{align}
From the new starting point,
we then adiabatically increase the drive amplitude from zero to its target value to arrive at the coupling-first (CF) dressed state $|\overline{i_q, j_s}\rangle \to|\overline{i_q, j_s}[\mathcal{E}_q]\rangle_{\rm CF}$ with corresponding quasi-energy $\epsilon_{\overline{i_q,j_s}}[\mathcal{E}_q]_{\rm CF}$. 

Thus, mimicking the Floquet branch analysis discussed in \cref{sec:nutshell}, the coupling-first Floquet branch analysis first considers all qubit-spectator couplings at their target value $J_{ij}$ before adding photons to the system. We can then succinctly summarize the procedure as
\begin{align}\label{eq:Floquet_BA}
    |i_q, j_s\rangle 
    \overset{\lambda}{\rightarrow} 
    |\overline{i_q, j_s}\rangle 
    \overset{\mathcal{E}_q}{\rightarrow} 
    |\overline{i_q,j_s}[\mathcal{E}_q]\rangle_{\rm CF},
\end{align}
where the arrows indicate labeling of the states by an adiabatic increase of the appropriate parameter, see \cref{fig:fig2}.
 
\subsection{Drive-first Floquet branch analysis}

In contrast to the coupling-first Floquet branch analysis, the drive-first Floquet approach begins by increasing the drive amplitude $\mathcal{E}_q$ in the uncoupled system, i.e $\lambda = 0$. As discussed in \cref{sec:nutshell}, by starting in the state $|i_q, j_s\rangle$ and adiabatically increasing the drive strength $\mathcal{E}_q$, we obtain a drive-first (DF) Floquet mode $|i_q[\mathcal{E}_q], j_s \rangle$ and quasi-energy $\epsilon_{i_q,j_s}[\mathcal{E}_q]$ of $\hat{H}_{\lambda = 0, \mathcal{E}_q}(t)$. These modes are necessarily tensor products of the qubit and spectator degrees of freedom. Starting from such an unentangled state, we adiabatically increase the qubit-spectator coupling $\lambda = 0 \to 1$. This results in a Floquet mode $|\overline{i_q, j_s} [\mathcal{E}_q] \rangle_{\rm DF}$ and corresponding quasi-energy $\epsilon_{\overline{i_q,j_s}}[\mathcal{E}_q]_{\rm DF}$ of the target Hamiltonian $\hat{H}_{\lambda = 1, \mathcal{E}_q}(t)$. 

This procedure mimics alternate labeling methods to the branch analysis presented in literature \cite{Nesterov_MIST_fluxonium} , wherein the system under consideration begins at its target photon number before a coupling parameter is introduced. The construction of the Floquet spectators branches can be succinctly written as
\begin{equation}\label{eq:Floquet_SA}
    |i_q, j_s\rangle 
    \overset{\mathcal{E}_q}{\rightarrow} 
    |i_q[\mathcal{E}_q], j_s \rangle 
    \overset{\lambda}{\rightarrow} 
    |\overline{i_q, j_s} [\mathcal{E}_q] \rangle_{\rm DF},
\end{equation}
where the arrows again indicate labeling of the states by an adiabatic increase of the appropriate parameter, see \cref{fig:fig2}.

\subsection{Spectator induced critical photon number}

With these two labeling methods at our disposal, we now have a straightforward way to identify at what photon number the spectators induce unwanted transitions outside the computational subspace. This is because, the two procedures may not give the same result, i.e., $|\overline{i_q, j_s} [\mathcal{E}_q] \rangle_{\rm CF}$ and $|\overline{i_q, j_s} [\mathcal{E}_q] \rangle_{\rm DF}$ need not coincide. The physical reason is simple: increasing the coupling and then adding photons is not the same as adding photons and then increasing the coupling. Note, however, that the collection of all states must be the same $\{|\overline{i_q, j_s} [\mathcal{E}_q] \rangle_{\rm CF}\}_{i_q, j_s} = \{|\overline{i_q, j_s} [\mathcal{E}_q] \rangle_{\rm DF}\}_{i_q, j_s}$ since they are the Floquet modes of the same Hamiltonian $\hat{H}_{\lambda = 1, \mathcal{E}_q}(t)$. 

We now argue that the resulting individual states disagree \textit{precisely} when the spectators induce an avoided crossing in the quasienergy spectrum that would otherwise be absent. By adding photons with the spectators already coupled to the qubit, the state $|\overline{i_q, j_s} [\mathcal{E}_q] \rangle_{\rm CF}$ follows the branch adiabatically through any avoided crossing encountered along the photon-number ramp. In contrast, since the coupling to the spectators is introduced after the target photon number has been reached, the state $|\overline{i_q, j_s} [\mathcal{E}_q] \rangle_{\rm DF}$ is unaffected by avoided crossings that occur at lower photon numbers. In this sense, the drive-first Floquet branch analysis follows the diabatic branches at these crossings. Further, since the latter captures all avoided crossings that arise solely due to the qubit, any discrepancy in the state assignment must be due to a photon-induced avoided crossing that is only present due to the spectators.

As a result, we have a simple way to define spectator-induced critical photon number for a given bare state $|i_q, j_s\rangle$: it is the lowest photon number at which the two labeling methods disagree
\begin{align}\label{eq:n_crit_criteria}
    n_{\rm crit}^{\rm spec}(i_q,j_s)
    =
    \underset{\mathcal{E}_q}{\min}
    \left(
    |\overline{i_q,j_s}[\mathcal{E}_q]\rangle_{\rm CF} 
    \neq
    |\overline{i_q,j_s}[\mathcal{E}_q]\rangle_{\rm DF}
    \right).
\end{align}
Because of the unbounded nature of the Hamiltonians we are working with, there is an infinite number of quasienergies gaps with arbitrary small sizes \cite{Hone1997gap}. To avoid capturing irrelevant anti crossings, we choose a fix increment step $\delta\mathcal{E}_q$ to build the branches. The spectator-induced critical photon number therefore depends on this choice of increment. Following Ref.~\cite{Dumas2024}, throughout this article we fix $\delta\mathcal{E}_q/2\pi = 10 \: \text{MHz}$.

\subsection{Non-uniform spectator couplings $\lambda_{ij}$}
\label{sec:Multimode_branch_analysis,D}

By choosing a single scaling parameter $\lambda$ in~\cref{eq:H_multimode}, we are effectively only concerned with whether any spectator leads to a measurement-induced transition. If one is instead interested in determining if a specific spectator (or collection of spectators) is the root cause of drive-induced transitions, one can simply introduce a scaling parameter $\lambda_{ij}$ for each coupling $J_{ij} \to \lambda_{ij} J_{ij}$. The labeling is then redone by changing the order in which the couplings and drive are increased. Comparing the final states for these different procedures gives us information about which spectator induces this additional leakage. The logic is the same as in the previous subsection: changing the order in which the various $\lambda_{ij}$ are increased changes the location and nature of the avoided crossings.

For concreteness, consider the simplest non-trivial example: two uncoupled spectators with states labeled $|j_s\rangle \equiv |j_{s_1}, j_{s_2}\rangle$. Assume we tune the detunings and coupling strengths such that the first spectator induces leakage of the qubit, whereas the second spectator only trivially dresses the qubit. By construction, adiabatically increasing $\lambda_2$ never causes MIST, whereas $\lambda_1$ does. In other words, increasing the couplings $\lambda_1$ and $\lambda_2$ separately reveals which spectator induced the MIST of the qubit. This information is completely lost when increasing one global parameter $\lambda$ as in the previous subsection.

\section{Static coupling}
\label{sec:single_spectator}

We begin by considering a simple case for which our formalism is applicable: two capacitively-coupled transmons. One of these transmons, referred to as the qubit~($q$), is assumed to be coupled to a readout resonator. The other transmon will be referred to as the spectator~($s$). In the semi-classical approximation~\cite{Dumas2024}, the system Hamiltonian  takes the form 
\begin{equation}
    \hH(t) = \hH_q + \hH_s + J_{qs}\hn_q\hn_s + \mathcal{E}_q\cos(\omega_d t)\hn_q,
    \label{eq:qubit+spec}
\end{equation}
where $\hH_q$ and $\hH_s$ are the transmon Hamiltonians for the qubit and spectator, respectively, $J_{qs}$ denotes the qubit-spectator coupling strength, while $\mathcal{E}_q$ is the effective drive strength on the qubit. Throughout this section, we set $E_{C_q}/2\pi = 195 \: \text{MHz}$ and $E_{J_q}/E_{C_q} = 85$. These parameters correspond to a qubit frequency $\omega_q/2\pi = 4.88 \: \text{GHz}$ and an anharmonicity of $\alpha_q/2\pi = -216 \: \text{MHz}$. Additionally, we take the coupling strength between the qubit and the readout resonator to be $g/2\pi = 160 \: \text{MHz}$ and the drive frequency $\omega_d/2\pi = 7.5$ \: \text{GHz}.

Our ultimate goal is to vary both the coupling strength $g_{qs}$ and the spectator parameters to assess how deleterious multi-photon transitions impact the critical photon number. However, it is helpful to first perform the coupling-first and drive-first Floquet branch analyses on a fixed set of parameters to compare the two methods, build intuition, and to confirm that the observed avoided crossings and branch swapping have an impact on the dynamics of the system during readout.

\subsection{MIST induced by the spectator} 
In this subsection, we fix the spectator parameters $E_{C_s}/2\pi = 190 \: \text{MHz}$ and $E_{J_s}/2\pi = 16.1 \: \text{GHz}$, leading to a spectator frequency $\omega_s/2\pi = 4.75 \: \text{GHz}$ and an anharmonicity $\alpha_s/2\pi = -200 \: \text{MHz}$. Further, we fix the coupling strength $g_{qs}/2\pi = 5 \: \text{MHz}$. 

To compare the coupling-first and drive-first labeling methods, we track the average qubit population
\begin{equation}
    \langle\hat{N}_q \rangle_{\overline{i_q,j_s}} 
    \equiv 
    \sum_{i^{'}_q,j^{'}_s}i^{'}_q\left|\langle i^{'}_q,j^{'}_s| \overline{i_q,j_s}[\mathcal{E}_q]\rangle\right|^2,
    \label{eq:qubit_population}
\end{equation}
and spectator population
\begin{equation}
    \langle\hat{N}_s \rangle_{\overline{i_q,j_s}} 
    \equiv 
    \sum_{i^{'}_q,j^{'}_s}j^{'}_s\left|\langle i^{'}_q,j^{'}_s| \overline{i_q,j_s}[\mathcal{E}_q]\rangle\right|^2.
\end{equation}
as a function of the drive amplitude $\mathcal{E}_q$, or, equivalently, the average number of photons in the resonator \(\bar{n}_r\). The above expressions are the natural extension of the single-mode period-averaged transmon population of~\cref{eq:transmon_population}. As we stressed in ~\cref{sec:Multimode_branch_analysis}, it is crucial to note that for a fixed drive amplitude $\mathcal{E}_q$, the collection of all such populations is the same regardless of the labeling method. The two schemes simply assign labels to every Floquet mode. However, as noted in the previous subsection, the labels that the two methods assign to the Floquet modes will diverge at spectator-induced avoided crossings, leading to the the average populations to similarly diverge, i.e., $\langle\hat{N}_{q/s} \rangle_{\overline{i_q,j_s}_{\rm{CF}}} \neq \langle\hat{N}_{q/s} \rangle_{\overline{i_q,j_s}_{\rm{DF}}}$

\begin{figure}[ht!]
    \centering    \includegraphics[width=1\linewidth]{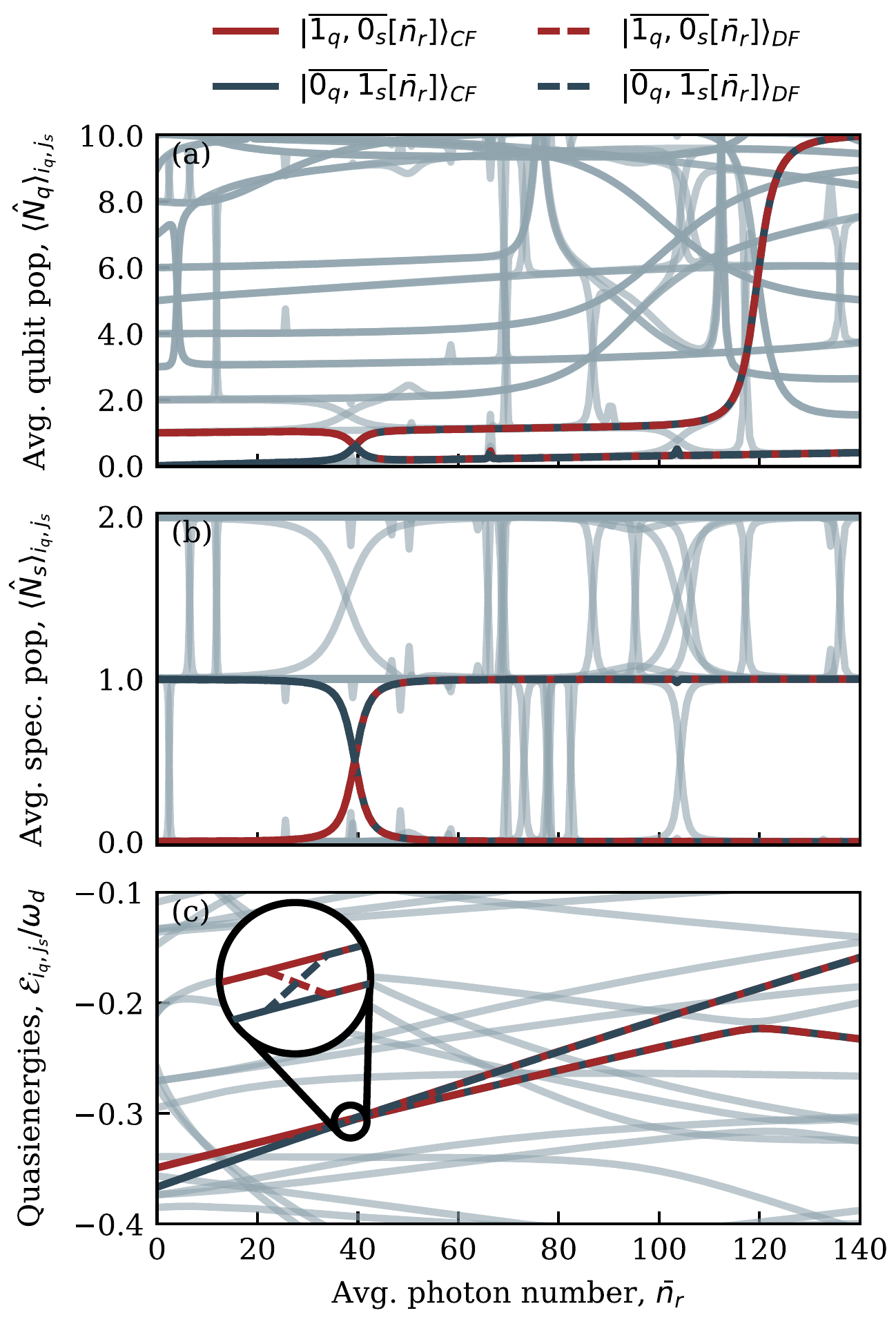}
    \caption{Example of spectator-induced MIST. (a) Average qubit  $\langle\hat{N}_q\rangle_{i_q,j_s}$ and (b) spectator $\langle\hat{N}_s\rangle_{i_q,j_s}$ population of the Floquet branches and (c) their corresponding quasieneriges as a function of the average resonator photon number $\bar{n}_r$. The populations and quasienergies of the $|\overline{1_q,0_s}[\bar{n}_r]\rangle$ and $|\overline{0_q,1_s}[\bar{n}_r]\rangle$ branches are highlighted in red and blue, respectively. The branches are labeled using either the coupling-first (solid lines) or drive-first (dashed lines) Floquet branch analysis methods. In (c), the quasienergies are defined with respect to the ground state energy of the system and subsequently folded into the interval $[-\omega_d/2,\omega_d/2]$. We observe a swapping of the average qubit and spectator populations of the coupler-first branches at $\bar{n}_r \approx 40$, leading to an excitation (relaxation) of the spectator conditioned on the qubit being in the first excited (ground) state. This results from an avoided crossing (subset) seen in coupling-first branches in the quasienergy spectrum (full lines). Importantly, a crossing is observed in the drive-first branches (dashed lines). The difference between the two branch analysis methods indicates a transition induced by the spectator.}
    \label{fig:qubit+spec}
\end{figure}

We confirm the differences between the two labeling methods in~\cref{fig:qubit+spec}(a)-(b), where we plot both the average populations of the qubit and spectator, respectively, as a function of the average photon number in the resonator. At sufficiently low drive amplitudes, the coupling-first and drive-first labeling methods give the same Floquet mode labels. However, near $\bar{n}_r \approx 40$ where a spectator-induced resonance occurs, the schemes disagree. The coupling-first Floquet state $|\overline{1_q,0_s}[\bar{n}_r]\rangle_{\rm CF}$ (solid red line) swaps its population with the Floquet state $|\overline{0_q,1_s}[\bar{n}_r]\rangle_{\rm CF}$ (solid blue line), as confirmed by comparing the two averaged populations at the point where the populations swap. In contrast, the drive-first Floquet state (dashed lines) continues to have, on average, a single excitation of the qubit and no excitation of the spectator. We confirm that this is due to an avoided crossing between the quasienergies of the Floquet modes involved in the swapping, as shown in~\cref{fig:qubit+spec}(c). We also stress that this behavior is only observed at spectator-induced avoided crossings. For instance, at a photon number of $\bar{n}_r \approx 120$, we see in~\cref{fig:qubit+spec}(a) that the drive-first Floquet branch undergoes another branch swapping. Although this is due to an avoided crossing and the state does indeed follow the coupling-first branch, this multi-photon transition does not involve the spectator, which effectively remains in its first excited state $\langle N_s\rangle_{\overline{1_q,0_s}} \approx$ 1.

While the branch analysis can pinpoint the number of photons at which a multi-photon resonance that causes a MIST event to occur, it does not predict the expected population exchange when going through that point during readout. This is effectively a Landau-Zener process, and the population transfer is therefore controlled by the magnitude of the anticrossing and the speed at which the resonance is traversed~\cite{Grifoni1998,Shillito2022,Dumas2024,Wang2025exp}. To explore this transfer of population in the presence of the spectator, we now investigate the coupled qubit-spectator dynamics. Recall that in~\cref{eq:qubit+spec}, we assumed that the drive amplitude $\mathcal{E}_q$ to be constant in time, which enabled our use of the Floquet theory. However, this effective off-resonant drive experienced by the qubit is a result of populating the cavity with a certain number of photons and is therefore time-dependent. Assuming a simple flat pulse, the drive amplitude then takes the form~\cite{Dumas2024}
\begin{equation}
\label{eq:drive on qubit}
\mathcal{E}_q(t) = 2g\sqrt{\bar{n}_r(t)},
\end{equation}
where 
\begin{equation}
    \bar{n}_r(t) = \bar{n}_{\infty}(1-e^{-\kappa t/2})^2
    \label{eq:td_amplitude}
\end{equation}
is the average photon number in the resonator, and $\overline{n}_{\infty}$ is the steady-state photon number. Here, we set $\kappa/2\pi = 7.95 \: \text{MHz}$ and vary $\overline{n}_\infty$ between 30 and 80.

\begin{figure}[t]
    \centering
    \includegraphics[width=1\linewidth]{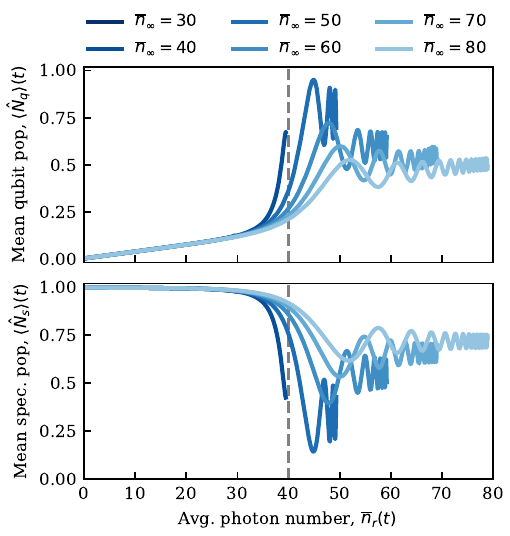}
    \caption{The time evolution of the (a) average qubit and (b) spectator population as a function of the resonator mean photon number, with the system initialized in the state $|\Psi(0)\rangle = |\overline{0_q,1_s}\rangle$. The dynamics is computed for six values of the steady-state photon number $\bar{n}_{\infty}$ ranging between 30 and 80 (see legend). The vertical dashed line indicates the critical photon number obtained via the multimode branch analysis method. The steady-state photon number $\overline{n}_\infty$ determines the speed of passage through the observed avoided crossing in the quasienergy spectrum at \(\bar{n}_{r} \approx 40\) photons (vertical dashed line) that is associated with the observed branch population swapping (see~\cref{fig:qubit+spec}). Larger $\overline{n}_\infty$ leads to a faster crossing through the resonance and therefore less population being transferred between the relevant coupling-first Floquet branches.}
    \label{fig:TD}
\end{figure}

Taking the initial state to be $|\Psi(0)\rangle = |\overline{0_q,1_s}\rangle$, we numerically solve the Schrödinger equation with the Hamiltonian given in~\cref{eq:qubit+spec}, thereby ignoring dissipation except for the finite filling rate $\kappa$ of the cavity. To characterize the resulting time-dependent state, we once again consider the average qubit $\hat{N}_q \equiv \sum_{i_q,j_s} i_q |i_q, j_s\rangle \langle i_q, j_s|$ and spectator $\hat{N}_s \equiv \sum_{i_q,j_s} j_s |i_q, j_s\rangle \langle i_q, j_s|$ populations. As shown in~\cref{fig:TD}, a parametric plot of these averages as a function of the drive amplitude confirms that, as predicted by our method, a spectator-induced state transition occurs at around $\bar{n}_r \approx 40$ (dashed vertical line). Note that as we increase the drive amplitude $\propto \bar{n}_{\infty}$, the speed at which the resonance is crossed increases. As seen in ~\cref{fig:TD} and in agreement with Landau-Zener theory~\cite{Grifoni1998}, the probability with which the population is transferred to the drive-first branch increases.

\subsection{MIST as a function of detuning and coupling strength}
\begin{figure*}[ht!]
    \centering
    \includegraphics[width=\textwidth]{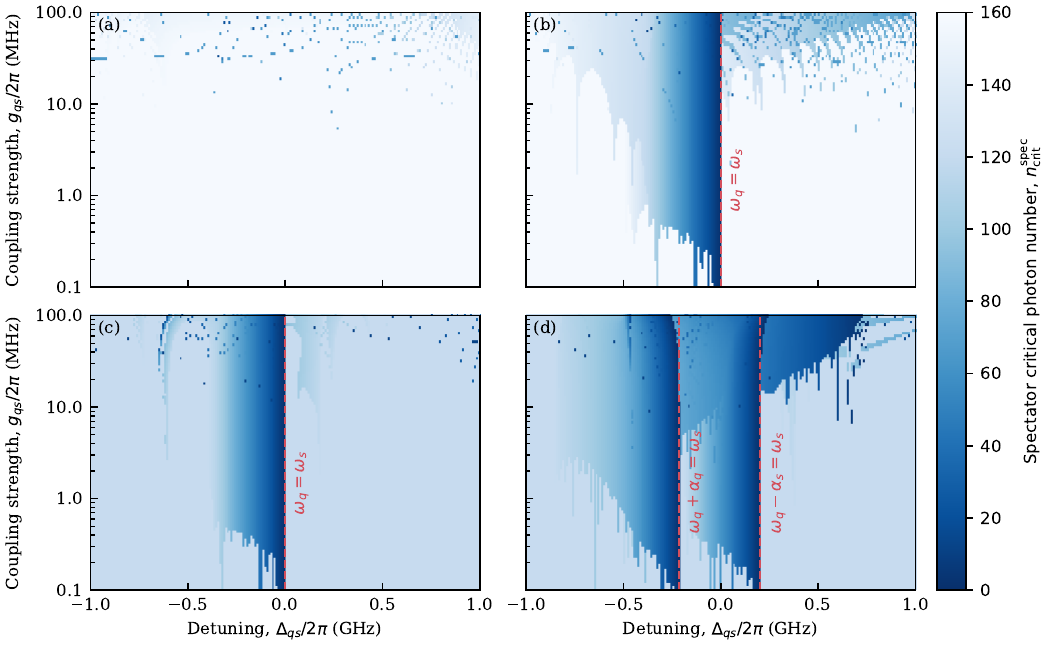}
    \caption{Critical photon numbers for the four computational states of the qubit and spectator as a function of the detuning $\Delta_{qs} = \omega_s-\omega_q$ and the  coupling strength between the qubit and the spectator, at $n_g = 0$. As discussed in the text, the critical photon numbers are extracted from both the average qubit and spectator populations.  (a) $|\overline{0_q,0_s}\rangle$. In the absence of a qubit or spectator excitation, the critical photon number is nearly uniform at $n_{crit} = 158$ and results from a multi-photon resonance between states $|0_q\rangle$ and $|14_q\rangle$ (see \cref{appendix:single mode ionization}). (b) $|\overline{0_q,1_s}\rangle$. In the presence of a spectator excitation, the background remains at $n_{crit} = 158$ and results from the same resonance as in panel (a). The large drop in $n_{crit}$ that is observed on top of the background (dark blue) results from the presence of a single-photon resonance at $\omega_q = \omega_s$ (red dashed line) leading to a swap between $|\overline{0_q,1_s}[\bar{n}_r]\rangle_{\rm CF}$ and $|\overline{1_q,0_s}[\bar{n}_r]\rangle_{\rm CF}$. Crucially, the impact of this resonance is extends to detunings as large as several hundreds of megahertz at typical couplings strengths.  (c) $|\overline{1_q,0_s}\rangle$. The overall background of $n_{crit} = 120$ results from a multi-photon resonance between $|1_q\rangle$ and $|12_q\rangle$. A large region of low  $n_{crit}$ is again observed in the vicinity of $\omega_q=\omega_s$. (d) $|\overline{1_q,1_s}\rangle$. The background of $n_{crit} = 120$ results from the same resonance as in (c). For this doubly-excited state, we now observe two regions with low $n_{crit}$ associated to a resonance between $|\overline{1_q,1_s}\rangle$ and $|\overline{0_q,2_s}[\bar{n}_r]\rangle_{\rm CF}$ at $\omega_q-\alpha_s=\omega_s$ (right red line) and between $|\overline{1_q,1_s}\rangle$ and $|\overline{2_q,0_s}[\bar{n}_r]\rangle_{\rm CF}$ at $\omega_q+\alpha_q=\omega_s$ (left red line).  
    } 
    \label{fig:n_crit}
\end{figure*}

We now extract the critical photon numbers, or equivalently the critical drive amplitudes, associated with the onset of unwanted qubit–spectator state transitions over a broad range of qubit–spectator detunings and coupling strengths. We keep the qubit and drive frequency fixed, taking the same parameters throughout this section. Keeping the anharmonicity of the spectator fixed at $\alpha_s / 2\pi = -200$~MHz, we vary its frequency in the range $ \omega_s /2 \pi \in [3.88, 5.88]$ GHz to ensure that the detuning $\Delta_{qs} \equiv \omega_s - \omega_q$ between the spectator and the qubit is in the range $\Delta_{qs} / 2\pi \in [-1 , 1]$~GHz. Finally, we vary the qubit-spectator coupling $g_{\rm qs}/2\pi \in [10^{-1}, 10^2]$~MHz.

The four computational states of the qubit spectator pair $|\overline{i_q, j_s}\rangle$ with $i_q, j_s = 0, 1$ are defined by starting with the corresponding bare states and adiabatically increasing the coupling $\lambda = 0 \to 1$. This
is precisely the first step of the coupling-first Floquet branch construction. For each of these states, we plot in \cref{eq:n_crit_criteria} the critical photon number (defined in~\cref{eq:n_crit_criteria}) as a function of the qubit-spectator detuning and coupling strength. We observe that the joint ground state $|\overline{0_q, 0_s}\rangle$ is largely unaffected by the presence of the spectator even at near-zero detuning, see panel~\ref{fig:n_crit}(a). This is merely the consequence of the small coupling strengths $g_{qs}$  that we consider, combined with the near excitation-preserving nature of the capacitive coupling between the qubit and spectator. In contrast, there are large regions of detuning where the critical photon number of both single-excitation states $|\overline{0_q, 1_s}\rangle$ and $|\overline{1_q, 0_s}\rangle$ is significantly reduced, see panels~\ref{fig:n_crit}(b) and (c). As expected, the features occur near the point when the qubit is nearly resonant with the spectator, where the effective drive on the qubit enables the single-excitation exchange interaction between the two transmons.

The asymmetry between the positive and negative detuning observed in panels (b) and (c) can also readily be explained. Recall that when the transmon-drive detuning is negative $\Delta_{qd} = \omega_q - \omega_d < 0$, the ground state has a larger ac-Stark shift per photon than the first excited state,  $\chi_{0_q} > \chi_{1_q} >0$  \cite{Dumas2024}. Since the spectator is not driven directly, its effect on the ac-Stark shift is small and we have $\chi_{\overline{0_q,1_s}} > \chi_{\overline{1_q,0_s}} >0$, which means $|\overline{0_q, 1_s} \rangle$ has a larger pull than $|\overline{1_q, 0_s} \rangle$. Placing the qubit frequency below the spectator's $\Delta_{qs} < 0$ therefore ensures that the two Stark-shifted frequencies will eventually collide, with the $|\overline{0_q, 1_s} \rangle$ state eventually ``catching up" the $|\overline{1_q, 0_s} \rangle$ state in energy. This is however not a problem in the opposite case $\Delta_{qs} > 0$. This also underlies why the critical photon numbers increase as we move away from exactly zero detuning: it takes more photons to make this transition resonant.

The same reasoning applies to the two-excitation computational state $|\overline{1_q, 1_s}\rangle$, see panel~\ref{fig:n_crit}(d). That state interacts most strongly with $|\overline{0_q, 2_s}\rangle$ and $|\overline{2_s, 0_q}\rangle$, which are also in the two-excitation manifold. In contrast to the single-excitation manifold, the resonance conditions that may lead to MIST are shifted by the qubit anharmonicity $\alpha_q/2\pi = - 216$~MHz or the spectator anharmonicity $\alpha_s/2\pi = -200$~MHz. For a wide range of coupling strengths, we again observe relatively large regions of detuning where transitions can be induced by the measurement. Unlike the transitions between states in the single-excitation manifold, these can lead to leakage of the qubit or the spectator out of the computational subspace. Such errors can have a detrimental impact on the logical performance of quantum error correction stabilizer codes, leading to a need to use leakage-removal operations to restore the performance of these codes~\cite{Aliferis07, Fowler13, Ghosh13b, Ghosh15, Kelly15, Suchara15, Varbanov2020QEC, McEwen2021, Miao2022, Marshall2025Incoherent}.

Crucially, in the single- and two-excitation subspace, a reduction in the critical photon number is observed over a large range of qubit-spectator detuning. This is particularly challenging to avoid when using microwave-activated cross-resonance gates, for which the detuning between neighboring qubits is required to be smaller than the anharmonicity to enable a fast two-qubit interaction. More broadly, mitigating spectator-induced MIST may become increasingly difficult as fixed-frequency qubit processors continue to scale up, further compounding the issue of frequency crowding.  In principle, it is possible to mitigate the spectator-induced MIST by employing tunable couplers. However, as we now show, this approach introduces its own set of challenges.

\section{Tunable coupling}
\label{sec:STTC}

We now consider the situation where the qubit-qubit interaction is mediated by a frequency-tunable transmon-type coupler. To cancel unwanted residual \(ZZ\) interactions, we account for a direct capacitance between the pair of qubit \cite{Yan2018TunableCoupler,Stehlik2021Gates}. Here, we focus on the simplest instance of this situation, comprising a single pair of transmons interacting via a transmon coupler and a drive on one of the transmons representing the measurement. We refer to the measured transmon as the qubit ($q$), the second transmon qubit as the spectator ($s$), and the transmon used for the coupling as the coupler ($c$). Within the semi-classical approximation discussed in Sec. \ref{sec:nutshell}, the Hamiltonian describing this system takes the form
\begin{equation}
    \begin{split}
        \hH(t) &= \sum_{i=q,c,s}\hH_i + \mathcal{E}_q\cos(\omega_dt)\hn_{q}\\ 
        &+ J_{qs}\hn_q\hn_s + J_{qc}\hn_q\hn_c + J_{sc}\hn_s\hn_c, 
        \label{eq:sttc}
    \end{split}
\end{equation}
where  $\hH_{i} = 4E_{C_{i}}\hn_{i} - E_{J_{i}}\cos\hat{\varphi}_{i}$ denotes the bare Hamiltonian of each subsystems. 

Throughout this section, we take the qubit and drive parameters to be the same as in the previous section. The other parameters, i.e., the spectator, coupler, and the coupling strengths $J_{ij}$, are chosen to be close to experimental values found in the literature \cite{Sung2021Gates}. More precisely, we fix the qubit-coupler and spectator-coupler coupling at $g_{qc}/2\pi = g_{sc}/2\pi = 70 \: \text{MHz}$, and the direct coupling between the qubit and the spectator to $g_{qs}/2\pi  = 5 \: \text{MHz}$. We fix the spectator parameters to the values used in \cref{sec:single_spectator}, namely a frequency of $\omega_{s}/2\pi = 4.75 \: \text{GHz}$ and anharmonicity $\alpha_{s}/2\pi = -200 \: \text{MHz}$. Moreover, the frequency of the coupler is chosen as $\omega_c/2\pi = 5.80 \: \text{GHz}$ and the anharmonicity $\alpha_c/2\pi = -90 \: \text{MHz}$.

As above, we characterize the mutual influence of the qubit, spectator, and coupler on MIST. One might expect an additional mode to worsen readout performance: additional modes imply more branches and more opportunities for collisions between the quasienergies. We first show that, surprisingly, there exists parameters that lead to a \textit{reduction} in spectator-induced leakage. 

\subsection{Impact of the coupler on the spectator's MIST}

\begin{figure}[t]
    \centering\includegraphics[width=\linewidth]{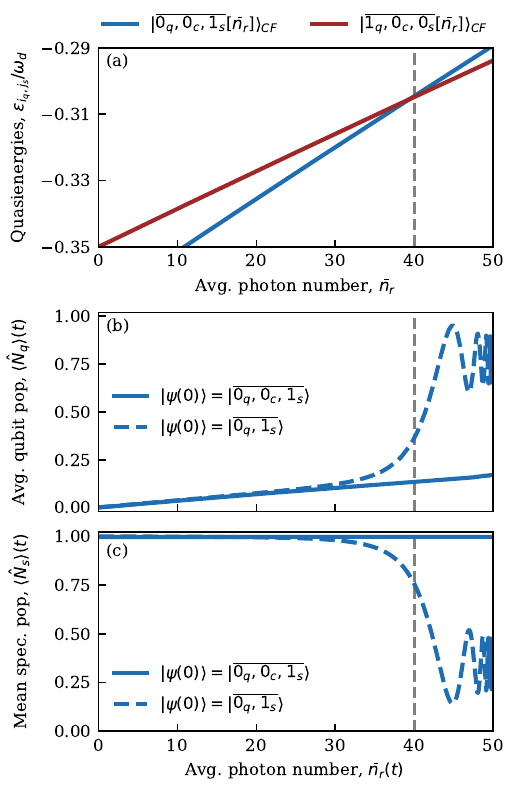}
\caption{Example of a spectator-induced MIST in the presence of a coupler mode. (a) Quasienergies of the coupling-first Floquet branches $|\overline{0_q,0_c,1_s}[\bar{n}_r]\rangle_{\rm CF}$ (blue) and $|\overline{1_q,0_c,0_s}[\bar{n}_r]\rangle_{\rm CF}$ (red). No anticrossing of the quasienergies is observed in the presence of the coupler. (b) Mean qubit population and (c) mean spectator population as a function of the mean photon number in the resonator. In the absence of the  coupler (dashed lines) we observe a large populations variation around the avoided crossing at $\bar{n}_r \sim 40$. In contrast, when the coupler is absent (full lines) no abrupt changes in the qubit population is observed.
The parameters are $g_{qc}/2\pi=g_{sc}/2\pi= 70 \: \text{MHz}$, $g_{qs}/2\pi = 5 \: \text{MHz}$ in the presence of the coupler and $g_{qc}=g_{sc}=0$ $g_{qs}/2\pi = 5 \: \text{MHz}$ in its absence.
}
    \label{fig:floquet_coupler}
\end{figure}

Following the previous section, we compute the coupling-first  
branch quasienergies $\epsilon_{\overline{i_q,k_c,j_s}}[\mathcal{E}_q]$ as a function of the average photon number \(\bar{n}_r\). Because the qubit and spectator parameters are unchanged from the previous section, we expect a drive-induced swap in the one-excitation subspace at $\bar{n}_r \sim 40$. In \Cref{fig:floquet_coupler} (a), we show the quasienergies in the presence of the coupler, which is in its ground state. In the absence of the coupler, we observed an avoided crossing of the coupling-first branches as in \cref{fig:qubit+spec} (c) at $\bar{n}_r \sim 40$ with the corresponding swap in the average populations, see \cref{fig:qubit+spec} (a,b). In the presence of the coupler, this avoided crossing becomes a simple crossing at the same photon number: it appears as though the coupler has canceled the spectator-induced MIST of the qubit.

To confirm this, in \cref{fig:floquet_coupler} we plot the average (b) qubit and (c) spectator populations as obtained from integrating the Schrödinger equation with the Hamiltonian \cref{eq:sttc} and the drive \cref{eq:drive on qubit} with $\overline{n}_\infty = 50$. The dashed lines are obtained in the absence of the coupler and show the expected rapid change in populations at the anticrossing (see also \cref{fig:TD} obtained for the same parameters with different $\bar n_\infty$). On the other hand, when the coupler is present, no rapid change of population is observed (full lines). Because the speed of passage through the resonance is the same with and without the coupler, we conclude that it is the smaller size of the quasienergy gap that modifies the Landau-Zener transition probabilities. This change is coupled-state dependent: taking the coupler to be in its first excited state, we again observe an abrupt change in populations at the anticrossing (not shown). This observation emphasizes the importance of accounting for the quantum state of multiple modes of the circuit.

\begin{figure*}[ht!]
    \centering
    \includegraphics[width=\textwidth]{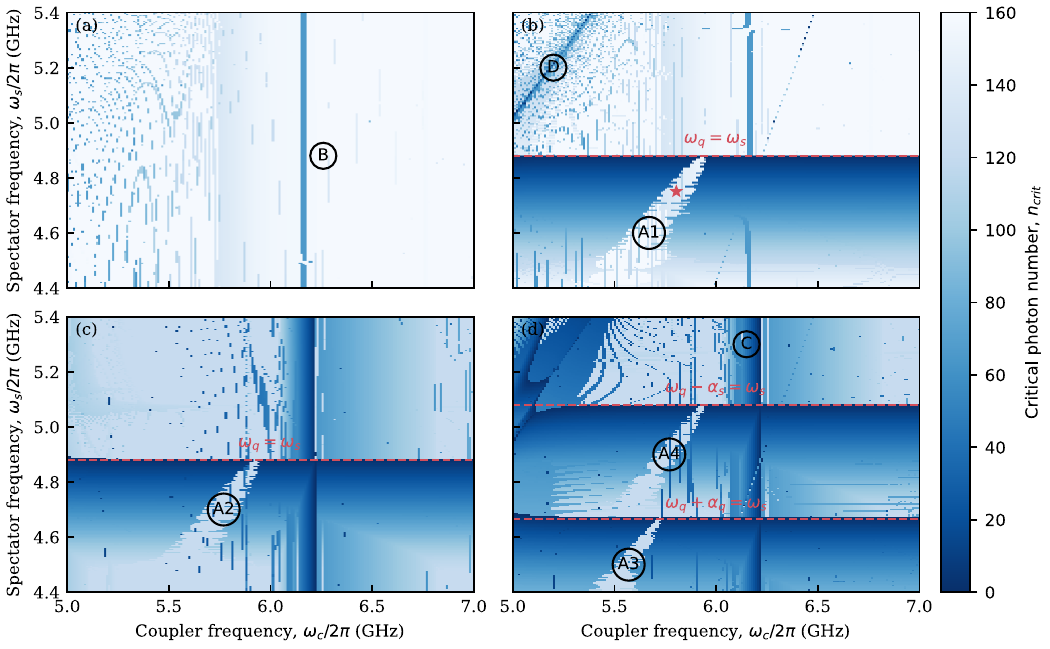}
    \caption{The extracted critical photon numbers as a function of the coupler frequency, \(\omega_{c}\) and spectator frequency \(\omega_{s}\)for each of the four computational states of the system: $\ket{\overline{0_q,0_c,0_s}}$ in (a), $\ket{\overline{0_q,0_c,1_s}}$ in (b), $\ket{\overline{1_q,0_c,0_s}}$ in (c) and $\ket{\overline{1_q,0_c,1_s}}$ in (d). Here, we consider no charge offset for each transmon. The extracted maximum critical photon number in (a, b) is $n_{crit} \approx 158$, while in (c, d) it is $n_{crit} \approx 120$, corresponding to transitions not involving any spectator or coupler states (see~\cref{fig:n_crit}). Resonances inducing state transitions involving the coupler appear as vertical features with a generally weak dependence on the spectator frequency. Similarly, transitions involving only the qubit and spectator states appear as horizontal features, with the frequency conditions of these resonances shown with solid red lines. However, we now also observe diagonal features, where both the coupler and spectator modes affect the observed MIST. Importantly, this can lead to a suppression of some transitions involving the spectator mode, thereby resulting in higher critical photon numbers. The red star in panel (b) indicates the parameter choice we considered for the text.}
    \label{fig:coupler}
\end{figure*}

\subsection{MIST as a function of detunings}

We now examine how the presence of the coupler affects the critical photon number over a range of coupler and spectator frequencies. \Cref{fig:coupler} shows, as a function of $\omega_s$ and $\omega_c$, the spectator-induced critical photon number for the computational states of the qubit and spectator, and no excitation in the coupler. Several notable features can be seen. 

First, for states containing at least one excitation, the critical photon number exhibits abrupt drops at specific spectator frequencies over a broad range of coupler frequencies. As indicated by the dashed red horizontal lines in panels (b), (c), and (d), these drops occur when the state of interest becomes resonant with another state within the same excitation manifold. These resonance frequencies are consistent with those identified in \cref{sec:single_spectator}.

Moreover, near certain one-excitation resonances, we observe extended diagonal regions with unusually large critical photon numbers for states containing at least one excitation. These features appear as diagonal bands in the $(\omega_s,\omega_c)$ plane, see e.g., the feature labeled \circled{A1} in panel (b). Because they occur in the vicinity of excitation-preserving resonances that otherwise produce strong reductions in the critical photon number, the enhanced critical photon number in these regions is attributed to a suppression of the matrix element responsible for the mixing between the states. Indeed, the charge drive on the qubit is the only term coupling the eigenstates of the static undriven Hamiltonian and, in the aforementioned regions, we numerically find that the spectator ground state and first excited state are uncoupled by the drive
\begin{equation}
    \langle \overline{i_q,0_c,0_s}|\hn_q|\overline{j_q,0_c,1_s}\rangle \approx 0
    \label{eq:num_cancellation01}
\end{equation}
for any qubit states $i_q$ and $j_q$. The same phenomenon is responsible for the diagonal regions with unusually large critical photon numbers labeled \circled{A2} and \circled{A3}. This cancellation can be understood from a Bogoliubov transformation of the coupler-spectator system. Indeed, in \cref{app:cancelation} we show that the matrix elements of $\hn_q$ connecting the spectator ground state to the spectator first excited state are canceled when
\begin{equation}
    \omega_c - \omega_s = g_{sc}\left(\dfrac{g_{qc}}{g_{qs}} -  \dfrac{g_{qs}}{g_{qc}}\right).
    \label{eq:cancel_01}
\end{equation}
For the parameters used in \cref{fig:coupler}, we expect the cancellation of the charge operator matrix element to occur for $(\omega_c - \omega_s)/2\pi = 0.975$ GHz with Eq. \eqref{eq:cancel_01}. Numerically, we find that this cancellation condition occurs is slightly shifted to   $(\omega_c - \omega_s)/2\pi = 1.06$ GHz, corresponding to the regions \circled{A1}, \circled{A2} and \circled{A3} (see \cref{app:cancelation}).

A similar matrix element cancellation also explains the feature labeled \circled{A4} in panel (d). In this region, we numerically find that, for any qubit states $i_q$ and $j_q$, we have 
\begin{equation}
    \langle \overline{i_q,0_c,1_s}|\hn_q|\overline{j_q,0_c,2_s}\rangle \approx 0.
    \label{eq:num_cancellation12}
\end{equation}
Because this matrix element involves up to two spectator excitations, we observe that this cancellation is now shifted by one spectator anharmonicity compared to \cref{eq:num_cancellation01}.
Importantly, we note that these regions where the critical photon number is large because of matrix elements cancellation do not coincide with the region where the $ZZ$ interaction is canceled, see \cref{app:cancelation} for details.

We also observe several vertical structures corresponding to a reduction of the critical photon number at approximately fixed coupler frequencies, e.g, see the feature labeled \circled{B} in panel (a). Given that these features are largely insensitive to the spectator frequency, they originate from coupler-specific processes. With the chosen parameter set, the process in question is a swap between the ground state of the qubit and its eighth excited state. This avoided crossing exists even without the presence of the spectator and coupler but, in the absence of these modes, the size of the gap is too small to be captured. 
This specific resonance is only revealed when the qubit is in its ground state, which explain why this feature is observed in panels (a) and (b) but not (c) or (d).

The region labeled \circled{C} in panel (d) corresponds to a single-photon resonance consisting of a swap between the first excited state of the qubit and the second excited state of the coupler. As such, this only appears when the qubit is in its first excited state, which explain why we also see this feature in panel (c) but not in (a,b).

Finally, the diagonal feature labeled \circled{D} in panel (b) corresponds to a resonance between the spectator and the coupler. This resonance is only a present when the spectator is in its first excited state, which explains why we also see this feature in panel (d) but not in (a) or (c).

\section{Conclusion and outlook}
\label{sec:conclusion}

We have introduced a framework to characterize measurement-induced state transitions induced by spectator modes. Our method is based on two different labeling schemes, the coupling-first Floquet branch and drive-first Floquet branch analysis. The two methods differ by the order in which the coupling to the spectator and the readout drive are taken into account when studying measurement-induced transitions. 
When the two labeling methods result in the same branch identification, it can be concluded that the spectator has no impact on measurement-induced transition of the qubit. On the other hand, obtaining different results indicates that the presence of the spectator modifies MIST of the qubit. As examples of the method, we explored MIST in the context of two transmons interacting via static or tunable coupling.

With static coupling, we observe that the effect of the spectator on the measurement-induced transition is concentrated around single-photon resonance conditions between the qubit and the resonator. Although the presence of these resonances are expected, we find that the region where the qubit is noticeably influenced by the spectator can span a wide range of frequencies near those resonances. Indeed, at typical qubit-qubit coupling strengths, the spectator-induced effects extend over frequency ranges of several hundred of megahertz around the resonance conditions. Such broad regions of influence can be particularly troublesome in the context of large-scale superconducting quantum processors where this can make it difficult to avoid all potentially harmful frequency configurations. 

By allowing the coupling to be turned off, the use of frequency-tunable couplers between qubits can, in principle, alleviate some of the above-mentioned frequency crowding issues. However, because couplers have internal modes that can impact measurement-induced transitions, their use comes with its own set of constraints on frequency placement. When these constraints are satisfied, the use of a coupler can help mitigate the impact of the spectator on MIST of the measured qubit. This mitigation can, however, be coupler-state dependent. Moreover, importantly the regions of parameter space where spectator effects can be significantly suppressed by the use of a coupler do not necessarily coincide with the regions where the residual ZZ interaction between the qubit and the spectator vanishes. In other words, optimizing the coupler so as to cancel the static ZZ coupling does not automatically minimize spectator-induced MIST. Overall, these results illustrate the importance of considering qubits within the full circuit when optimizing readout. Because ionization is not limited to readout but more generally occurs in strongly driven nonlinear circuits, the same conclusion can be expected to apply to other situations such as parametric couplers, reset, and state stabilization.

\section*{Acknowledgments}

This material is based upon work supported by the U.S. Department of Energy, Office of Science, National Quantum Information Science Research Centers, Quantum Systems Accelerator. Additional support is acknowledged from NSERC, the Ministère de l’Économie et de l’Innovation du Québec.

\appendix

\section{Single mode MIST}
\label{appendix:single mode ionization}

\begin{figure}[ht!]
    \centering
    \includegraphics[width=\linewidth]{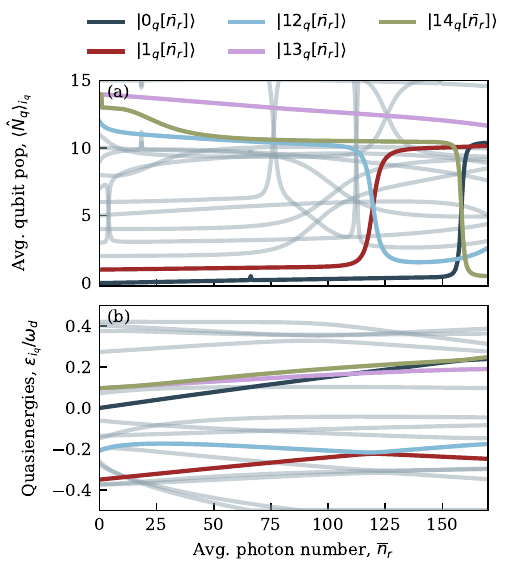}
    \caption{(a) Mean excitation number of the Floquet branches and (b) quasienergy spectrum as a function of the average photon number $\bar{n}_r$.}
    \label{fig:qubit}
\end{figure}

In the main text, we used the same qubit transmon for the two examples in Sec. \ref{sec:single_spectator} and Sec. \ref{sec:STTC}. Using the tools of Sec. \ref{sec:nutshell}, we characterize the MIST of the qubit alone. For this transmon, we fixed $E_C/2\pi = 195$ MHz and $E_J/E_C = 85$. This corresponds to a qubit frequency $\omega_q/2\pi = 4.88$ GHz anharmonicity $\alpha_q/2\pi = -215$ MHz. We consider the specific case where the gate charge is zero. The coupling between the qubit and the readout resonator is $g/2\pi = 160$ MHz, with a drive frequency of $\omega_d/2\pi = 7.5$ GHz.
In Fig. \ref{fig:qubit}, we show (a) the averaged transmon excitation for each Floquet branches and (b) the quasienergy spectrum as a function of the mean resonator photon number $\bar{n}_r$. The branch $|0_q[\bar{n}_r]\rangle$ swaps with the branch $|12_q[\bar{n}_r]\rangle$ around $\bar{n}_r = 160$ photons, and the branch $|1_q[\bar{n}_r]\rangle$ swaps with the branch $|14_q[\bar{n}_r]\rangle$ around $\bar{n}_r = 120$. These two swaps correspond to an avoided crossing in the quasienergy spectrum between the branches of interest. For the critical photon number, we use the same criteria as in \cite{Dumas2024}. It is defined as the minimum photon number $\bar{n}_r$ at which the average branch population reaches $\langle\hat{N_q}\rangle_{0_q} = 2$ for the ground state, and $\langle\hat{N_q}\rangle_{1_q} = 3$ for the excited state. Here we find that $\bar{n}_{\rm crit, 0_q} = 158$ and $\bar{n}_{\rm crit, 1_q} = 120$.

\section{Cancellation of the qubit-spectator induced MIST}\label{app:cancelation}

\begin{figure*}[ht!]
    \centering
    \includegraphics[width=\textwidth]{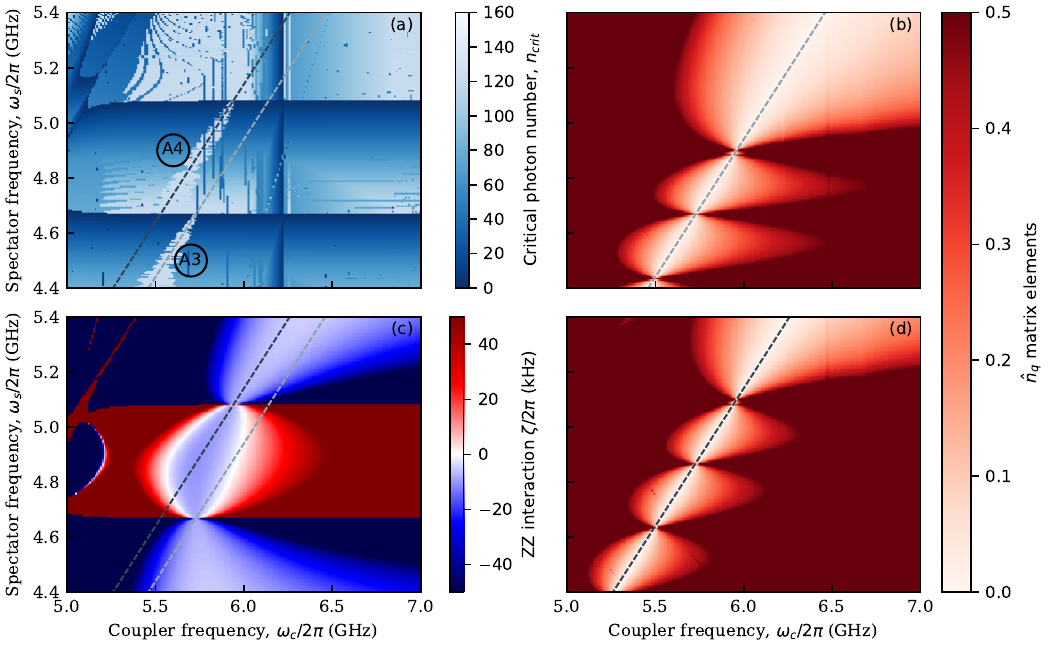}
    \caption{(a) Critical photon numbers of the state $\ket{\overline{1_q,0_c,1_s}}$ as a function of coupler \(\omega_{c}\) and spectator frequency \(\omega_{s}\). (b) Sum of the amplitude of every charge qubit matrix elements connecting the ground state of the spectator to its first excited state subspace, when the coupler is in the ground state, i.e., $\sum_{i_q,j_q} \left|\langle \overline{i_q,0_c,0_s} |\hat{n}_q|\overline{j_q,0_c,1_s} \rangle \right|$. This coupling is suppressed along the gray dashed diagonal line where $\omega_c - \omega_s \approx 1.06$ GHz. 
    The same gray dashed diagonal line is reported in (a) and (c). (c) ZZ interaction rate between the qubit and the spectator, defined as $\zeta = E_{\overline{1_q,0_c,1_s}} - E_{\overline{1_q,0_c,0_s}} - E_{\overline{0_q,0_c,1_s}} + E_{\overline{0_q,0_c,0_s}}$. (d) Sum of the amplitude of every charge qubit matrix elements connecting the first excited state of the spectator to its second excited state subspace, when the coupler is in the ground state, i.e., $\sum_{i_q,j_q} \left|\langle \overline{i_q,0_c,1_s} |\hat{n}_q|\overline{j_q,0_c,2_s} \rangle \right|$. This coupling is suppressed along the black diagonal line where $\omega_c - \omega_s \approx 0.86$ GHz. This diagonal line is also reported in (a) and (c). The frequency shift of the two diagonal lines implies that the large critical photon number does not occur at the zero-ZZ point.}
    \label{fig:matrix_element}
\end{figure*}

To understand the features \circled{A1}-\circled{A4} in \cref{fig:coupler}, we will first approximate the spectator and coupler as Kerr oscillators and diagonalize the linear part of the spectator-coupler Hamiltonian. Within the rotating-wave approximation, this can be achieved by defining rotated canonical spectator and coupler annihilation operators
\begin{align}
    & \hb_s'  = \cos(\Lambda)\hb_s - \sin(\Lambda)\hb_c \\
    & \hb_c' = \cos(\Lambda)\hb_c + \sin(\Lambda)\hb_s,
\end{align}
where $\Lambda = (1/2) \arctan(2\lambda)$ with $\lambda = g_{sc}/(\omega_c-\omega_s)$. Moreover, given that we focus on the one-excitation subspace of the spectator and coupler, we momentarily ignore their anharmonicities. The Hamiltonian expressed in this new basis and within these approximations take the form
\begin{equation}
    \begin{split}
        \hat{H}
        &=  \hat{H}_q + \hat{H}_{sc} \\
        &+  g_{qc}'(\hb_q^\dag + \hb_q)(\hb_c'^\dag + \hb_c') + g_{qs}'(\hb_q^\dag + \hb_q)(\hb_s'^\dag + \hb_s'),
        \label{eq:H+V}
    \end{split}
\end{equation}
where $\hH_{sc} = \omega_s'\hb_s'^\dag\hb_s + \omega_c'\hb_c'^\dag\hb_c$, with the new frequencies
\begin{align}
    & \tilde{\omega}_s = \dfrac{1}{2}\left(\omega_s + \omega_c - \sqrt{(\omega_c-\omega_s)^2 + 4g_{sc}^2}\right) \\
    & \tilde{\omega}_c = \dfrac{1}{2}\left(\omega_s + \omega_c + \sqrt{(\omega_c-\omega_s)^2 + 4g_{sc}^2}\right)
\end{align}
and the new coupling strength
\begin{align}
    & g_{qc}' = g_{qc}\cos(\Lambda) + g_{qs}\sin(\Lambda) \\
    & g_{qs}' = g_{qs}\cos(\Lambda) - g_{qc}\sin(\Lambda).
\end{align}
In this basis, there are two pathways connecting the qubit to the spectator: one resulting from their direct interaction ($g_{qs}\cos\Lambda$) and a second resulting from the hybridization of the coupler to the spectator ($g_{qc}\sin\Lambda$). Independently of the qubit parameters, the matrix elements \cref{eq:num_cancellation01} are canceled when these two pathways destructively interfere with each other such that $g_{qs}' = 0$.
With $\tan(\arctan(2\lambda)/2) = (\sqrt{1 + 4\lambda^2} -1)/2\lambda$, this cancellation occurs when
\begin{equation}
    \omega_c - \omega_s = g_{sc}\left(\dfrac{g_{qc}}{g_{qs}} -  \dfrac{g_{qs}}{g_{qc}}\right).
\end{equation}
For the parameters used in Sec. \ref{sec:STTC}, we get 
$(\omega_c - \omega_s)/2\pi = 0.975$ GHz. From the numerical 
solution, we find that the zero and one excitation subspace of the spectator are decoupled from $\hat{n}_q$ when $(\omega_c - \omega_s)/2\pi = 1.06$ GHz, see Fig. 
\cref{fig:matrix_element}(b). This condition is exactly what we observe in region \circled{$A_3$} of \cref{fig:matrix_element}(a).

In \cref{fig:matrix_element}(d) we report the matrix elements between the first and second excited states of the spectator. Here, we find that the two subspaces are decoupled from $\hat{n}_q$ when $(\omega_c - \omega_s)/2\pi = 0.86$ GHz, which is the same condition as before, now shifted by one anharmonicity of the spectator. This cancellation corresponds exactly to the region \circled{$A_4$} of \cref{fig:matrix_element}(a). In  \cref{fig:matrix_element}(c), we plot the ZZ interaction between the qubit and the spectator. We observe that the region where the ZZ interaction is canceled doesn't coincide with this matrix element suppression. However, for some choice of frequencies, we observe both a cancellation of the matrix elements responsible of the state transition and of the ZZ interaction. 

\section{Impact of the gate charge}
\label{appendix:impact of gate charge}

\begin{figure*}[ht!]
    \centering
    \includegraphics[width=\textwidth]{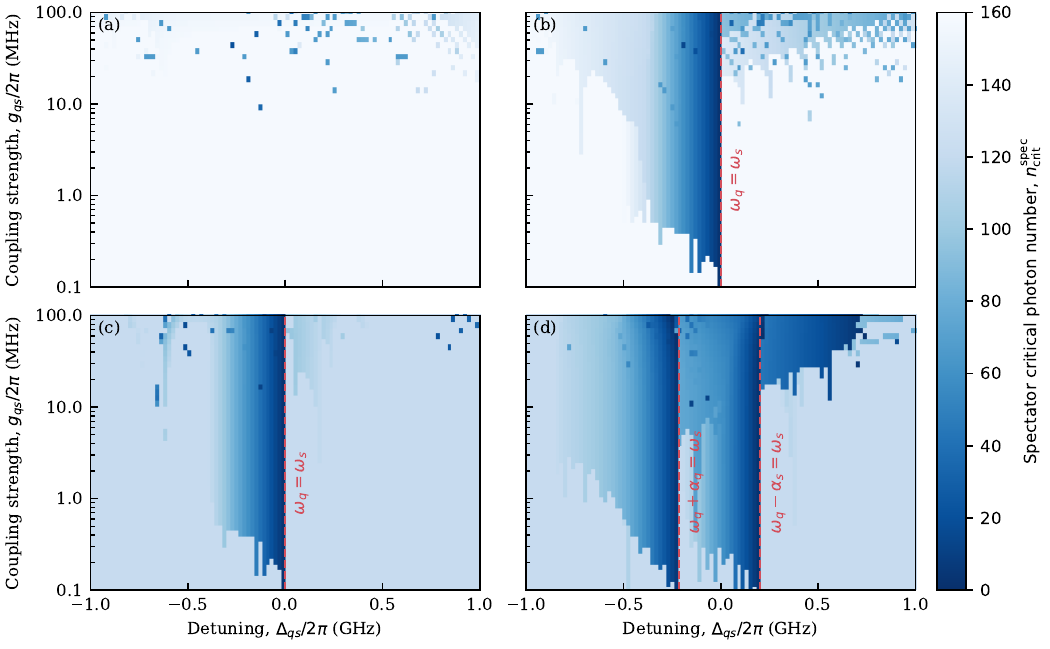}
    \caption{Critical photon number for the computational subspace, averaged over 50 values of gate charge for the qubit between 0 and 0.5. The spectator gate charge is fixed at 0. The x-axis represents the detuning $\Delta_{qs} = \omega_s-\omega_q$ and the y-axis represents the coupling strength between the qubit and the spectator.}
    \label{fig:ncrit_ng_qubit+spec+coup}
\end{figure*}

One important feature of the transmon is the small charge dispersion of the computational states. However, this charge dispersion rapidly increases for states near and above the well. As a results it is necessary to take the qubit gate charge $n_g$ into account~\cite{Dumas2024,Fechant2025exp}. The gate charge of the spectators mode are not relevant for two reasons. In Sec. \ref{sec:single_spectator} and Sec. \ref{sec:STTC} we have seen that an measurement-induced state transition can only occur between states which have differ at most by one spectator excitation. This is due to the weak coupling between the qubit and the spectator modes. Since deep in the transmon regime these state are are insensitive to the gate charge, spectator-induced MIST is also independent of the gate charge of the spectators.

To see the  this explicitly, we first account for the gate charge, in which case \cref{eq:qubit+spec} reads
\begin{equation}
    \hH(t) = \hH_q + \hH_s + J_{qs}\hn_q(\hn_s-n_{g_s}) + \mathcal{E}_q\cos(\omega_d t)\hn_q,
    \label{eq:qubit+spec+}
\end{equation}
where $\hH_s = 4E_{C_s}(\hn_s - n_{g_s})^2 - E_{J_s}\cos \hat{\varphi}_s$.
We fix the parameters to the same value as in \cref{sec:single_spectator} and average the critical photon number computed by sampling 50 values of the the spectator gate charge $n_{g_s}$ between 0 and 0.5. The results are presented in \cref{fig:ncrit_ng_qubit+spec+coup} where we  observe the spectator critical photon number as a function of the qubit-spectator detuning and the qubit-spectator coupling strength, averaged over the spectator gate charge. The figure is more sparsely sampled than \cref{fig:n_crit} due to the numerical cost of averaging over many values of the gate charge. We observe no significant differences with \cref{fig:n_crit} where the spectator gate charge was fixed at 0, showing the gate charge of the spectator has no impact on the spectator-induced MIST.

\newpage
\bibliography{references.bib}

@article{Verney2019,
  title = {Structural Instability of Driven Josephson Circuits Prevented by an Inductive Shunt},
  author = {Verney, Lucas and Lescanne, Rapha\"el and Devoret, Michel H. and Leghtas, Zaki and Mirrahimi, Mazyar},
  journal = {Phys. Rev. Appl.},
  volume = {11},
  issue = {2},
  pages = {024003},
  numpages = {12},
  year = {2019},
  month = {Feb},
  publisher = {American Physical Society},
  doi = {10.1103/PhysRevApplied.11.024003},
  url = {https://link.aps.org/doi/10.1103/PhysRevApplied.11.024003}
}

@article{Lescanne2019,
  title = {Escape of a Driven Quantum Josephson Circuit into Unconfined States},
  author = {Lescanne, Rapha\"el and Verney, Lucas and Ficheux, Quentin and Devoret, Michel H. and Huard, Benjamin and Mirrahimi, Mazyar and Leghtas, Zaki},
  journal = {Phys. Rev. Appl.},
  volume = {11},
  issue = {1},
  pages = {014030},
  numpages = {12},
  year = {2019},
  month = {Jan},
  publisher = {American Physical Society},
  doi = {10.1103/PhysRevApplied.11.014030},
  url = {https://link.aps.org/doi/10.1103/PhysRevApplied.11.014030}
}

@article{Thorbeck2024,
  title = {Readout-Induced Suppression and Enhancement of Superconducting Qubit Lifetimes},
  author = {Thorbeck, Ted and Xiao, Zhihao and Kamal, Archana and Govia, Luke C. G.},
  journal = {Phys. Rev. Lett.},
  volume = {132},
  issue = {9},
  pages = {090602},
  numpages = {6},
  year = {2024},
  month = {Feb},
  publisher = {American Physical Society},
  doi = {10.1103/PhysRevLett.132.090602},
  url = {https://link.aps.org/doi/10.1103/PhysRevLett.132.090602}
}

@article{Spring2025,
  title = {Fast Multiplexed Superconducting-Qubit Readout with Intrinsic Purcell Filtering Using a Multiconductor Transmission Line},
  author = {Spring, Peter A. and Milanovic, Luka and Sunada, Yoshiki and Wang, Shiyu and van Loo, Arjan F. and Tamate, Shuhei and Nakamura, Yasunobu},
  journal = {PRX Quantum},
  volume = {6},
  issue = {2},
  pages = {020345},
  numpages = {23},
  year = {2025},
  month = {Jun},
  publisher = {American Physical Society},
  doi = {10.1103/PRXQuantum.6.020345},
  url = {https://link.aps.org/doi/10.1103/PRXQuantum.6.020345}
}

@misc{Beaulieu2026,
      title={Fast, high-fidelity Transmon readout with intrinsic Purcell protection via nonperturbative cross-Kerr coupling}, 
      author={Guillaume Beaulieu and Jun-Zhe Chen and Marco Scigliuzzo and Othmane Benhayoune-Khadraoui and Alex A. Chapple and Peter A. Spring and Alexandre Blais and Pasquale Scarlino},
      year={2026},
      eprint={2601.04975},
      archivePrefix={arXiv},
      primaryClass={quant-ph},
      url={https://arxiv.org/abs/2601.04975}, 
}

@misc{Mori2025,
      title={Suppression of measurement-induced state transitions in $\cos{\phi}$-coupling transmon readout}, 
      author={Cyril Mori and Francesca D Esposito and Alexandru Petrescu and Lucas Ruela and Shelender Kumar and Vishnu Narayanan Suresh and Wael Ardati and Dorian Nicolas and Giulio Cappelli and Arpit Ranadive and Gwenael Le Gal and Martina Esposito and Quentin Ficheux and Nicolas Roch and Olivier Buisson},
      year={2025},
      eprint={2509.05126},
      archivePrefix={arXiv},
      primaryClass={quant-ph},
      url={https://arxiv.org/abs/2509.05126}, 
}

@article{Grifoni1998, 
year = {1998}, 
title = {{Driven quantum tunneling}}, 
author = {Grifoni, Milena and Hänggi, Peter}, 
journal = {Physics Reports}, 
issn = {0370-1573}, 
doi = {10.1016/s0370-1573(98)00022-2}, 
url = {http://www.sciencedirect.com/science/article/B6TVP-3V7WTNG-1/2/2abe8475f39c4989435da46dd8fe951b}, 
pages = {229 -- 354}, 
number = {5-6}, 
volume = {304}
}

@article{Breuer1989, 
year = {1989}, 
title = {{Quantum phases and Landau-Zener transitions in oscillating fields}}, 
author = {Breuer, H.P. and Holthaus, M.}, 
journal = {Physics Letters A}, 
issn = {0375-9601}, 
doi = {10.1016/0375-9601(89)90132-1}, 
pages = {507--512}, 
number = {9}, 
volume = {140}
}

@article{Blais2023rmp,
  title = {Circuit quantum electrodynamics},
  author = {Blais, Alexandre and Grimsmo, Arne L. and Girvin, S. M. and Wallraff, Andreas},
  journal = {Rev. Mod. Phys.},
  volume = {93},
  issue = {2},
  pages = {025005},
  numpages = {72},
  year = {2021},
  month = {May},
  publisher = {American Physical Society},
  doi = {10.1103/RevModPhys.93.025005},
  url = {https://link.aps.org/doi/10.1103/RevModPhys.93.025005}
}

@article{Koch2007transmon,
  title = {Charge-insensitive qubit design derived from the Cooper pair box},
  author = {Koch, Jens and Yu, Terri M. and Gambetta, Jay and Houck, A. A. and Schuster, D. I. and Majer, J. and Blais, Alexandre and Devoret, M. H. and Girvin, S. M. and Schoelkopf, R. J.},
  journal = {Phys. Rev. A},
  volume = {76},
  issue = {4},
  pages = {042319},
  numpages = {19},
  year = {2007},
  month = {Oct},
  publisher = {American Physical Society},
  doi = {10.1103/PhysRevA.76.042319},
  url = {https://link.aps.org/doi/10.1103/PhysRevA.76.042319}
}

@article{Walter2017readout,
  title = {Rapid High-Fidelity Single-Shot Dispersive Readout of Superconducting Qubits},
  author = {Walter, T. and Kurpiers, P. and Gasparinetti, S. and Magnard, P. and Poto\ifmmode \check{c}\else \v{c}\fi{}nik, A. and Salath\'e, Y. and Pechal, M. and Mondal, M. and Oppliger, M. and Eichler, C. and Wallraff, A.},
  journal = {Phys. Rev. Appl.},
  volume = {7},
  issue = {5},
  pages = {054020},
  numpages = {11},
  year = {2017},
  month = {May},
  publisher = {American Physical Society},
  doi = {10.1103/PhysRevApplied.7.054020},
  url = {https://link.aps.org/doi/10.1103/PhysRevApplied.7.054020}
}

@article{Sank2016mist,
  title = {Measurement-Induced State Transitions in a Superconducting Qubit: Beyond the Rotating Wave Approximation},
  author = {Sank, Daniel and Chen, Zijun and Khezri, Mostafa and Kelly, J. and Barends, R. and Campbell, B. and Chen, Y. and Chiaro, B. and Dunsworth, A. and Fowler, A. and Jeffrey, E. and Lucero, E. and Megrant, A. and Mutus, J. and Neeley, M. and Neill, C. and O'Malley, P. J. J. and Quintana, C. and Roushan, P. and Vainsencher, A. and White, T. and Wenner, J. and Korotkov, Alexander N. and Martinis, John M.},
  journal = {Phys. Rev. Lett.},
  volume = {117},
  issue = {19},
  pages = {190503},
  numpages = {6},
  year = {2016},
  month = {Nov},
  publisher = {American Physical Society},
  doi = {10.1103/PhysRevLett.117.190503},
  url = {https://link.aps.org/doi/10.1103/PhysRevLett.117.190503}
}

@article{Khezri2023mist,
  title = {Measurement-induced state transitions in a superconducting qubit: Within the rotating-wave approximation},
  author = {Khezri, Mostafa and Opremcak, Alex and Chen, Zijun and Miao, Kevin C. and McEwen, Matt and Bengtsson, Andreas and White, Theodore and Naaman, Ofer and Sank, Daniel and Korotkov, Alexander N. and Chen, Yu and Smelyanskiy, Vadim},
  journal = {Phys. Rev. Appl.},
  volume = {20},
  issue = {5},
  pages = {054008},
  numpages = {12},
  year = {2023},
  month = {Nov},
  publisher = {American Physical Society},
  doi = {10.1103/PhysRevApplied.20.054008},
  url = {https://link.aps.org/doi/10.1103/PhysRevApplied.20.054008}
}

@article{Swiadek2024readout,
  title = {Enhancing Dispersive Readout of Superconducting Qubits through Dynamic Control of the Dispersive Shift: Experiment and Theory},
  author = {Swiadek, Fran\ifmmode \mbox{\c{c}}\else \c{c}\fi{}ois and Shillito, Ross and Magnard, Paul and Remm, Ants and Hellings, Christoph and Lacroix, Nathan and Ficheux, Quentin and Zanuz, Dante Colao and Norris, Graham J. and Blais, Alexandre and Krinner, Sebastian and Wallraff, Andreas},
  journal = {PRX Quantum},
  volume = {5},
  issue = {4},
  pages = {040326},
  numpages = {14},
  year = {2024},
  month = {Nov},
  publisher = {American Physical Society},
  doi = {10.1103/PRXQuantum.5.040326},
  url = {https://link.aps.org/doi/10.1103/PRXQuantum.5.040326}
}

@article{Sunada2022readout,
  title = {Fast Readout and Reset of a Superconducting Qubit Coupled to a Resonator with an Intrinsic Purcell Filter},
  author = {Sunada, Y. and Kono, S. and Ilves, J. and Tamate, S. and Sugiyama, T. and Tabuchi, Y. and Nakamura, Y.},
  journal = {Phys. Rev. Appl.},
  volume = {17},
  issue = {4},
  pages = {044016},
  numpages = {12},
  year = {2022},
  month = {Apr},
  publisher = {American Physical Society},
  doi = {10.1103/PhysRevApplied.17.044016},
  url = {https://link.aps.org/doi/10.1103/PhysRevApplied.17.044016}
}

@article{Boissonneault2008,
  title = {Nonlinear dispersive regime of cavity QED: The dressed dephasing model},
  author = {Boissonneault, Maxime and Gambetta, J. M. and Blais, Alexandre},
  journal = {Phys. Rev. A},
  volume = {77},
  issue = {6},
  pages = {060305},
  numpages = {4},
  year = {2008},
  month = {Jun},
  publisher = {American Physical Society},
  doi = {10.1103/PhysRevA.77.060305},
  url = {https://link.aps.org/doi/10.1103/PhysRevA.77.060305}
}

@article{Boissonneault2009,
  title = {Dispersive regime of circuit QED: Photon-dependent qubit dephasing and relaxation rates},
  author = {Boissonneault, Maxime and Gambetta, J. M. and Blais, Alexandre},
  journal = {Phys. Rev. A},
  volume = {79},
  issue = {1},
  pages = {013819},
  numpages = {17},
  year = {2009},
  month = {Jan},
  publisher = {American Physical Society},
  doi = {10.1103/PhysRevA.79.013819},
  url = {https://link.aps.org/doi/10.1103/PhysRevA.79.013819}
}

@article{Petrescu2020,
  title = {Lifetime renormalization of driven weakly anharmonic superconducting qubits. II. The readout problem},
  author = {Petrescu, Alexandru and Malekakhlagh, Moein and T\"ureci, Hakan E.},
  journal = {Phys. Rev. B},
  volume = {101},
  issue = {13},
  pages = {134510},
  numpages = {27},
  year = {2020},
  month = {Apr},
  publisher = {American Physical Society},
  doi = {10.1103/PhysRevB.101.134510},
  url = {https://link.aps.org/doi/10.1103/PhysRevB.101.134510}
}

@article{Hanai2021,
  title = {Intrinsic mechanisms for drive-dependent Purcell decay in superconducting quantum circuits},
  author = {Hanai, Ryo and McDonald, Alexander and Clerk, Aashish},
  journal = {Phys. Rev. Res.},
  volume = {3},
  issue = {4},
  pages = {043228},
  numpages = {18},
  year = {2021},
  month = {Dec},
  publisher = {American Physical Society},
  doi = {10.1103/PhysRevResearch.3.043228},
  url = {https://link.aps.org/doi/10.1103/PhysRevResearch.3.043228}
}

@article{Shillito2022,
  title = {Dynamics of Transmon Ionization},
  author = {Shillito, Ross and Petrescu, Alexandru and Cohen, Joachim and Beall, Jackson and Hauru, Markus and Ganahl, Martin and Lewis, Adam G.M. and Vidal, Guifre and Blais, Alexandre},
  journal = {Phys. Rev. Appl.},
  volume = {18},
  issue = {3},
  pages = {034031},
  numpages = {11},
  year = {2022},
  month = {Sep},
  publisher = {American Physical Society},
  doi = {10.1103/PhysRevApplied.18.034031},
  url = {https://link.aps.org/doi/10.1103/PhysRevApplied.18.034031}
}

@article{Cohen2023,
  title = {Reminiscence of Classical Chaos in Driven Transmons},
  author = {Cohen, Joachim and Petrescu, Alexandru and Shillito, Ross and Blais, Alexandre},
  journal = {PRX Quantum},
  volume = {4},
  issue = {2},
  pages = {020312},
  numpages = {27},
  year = {2023},
  month = {Apr},
  publisher = {American Physical Society},
  doi = {10.1103/PRXQuantum.4.020312},
  url = {https://link.aps.org/doi/10.1103/PRXQuantum.4.020312}
}

@article{Li2023Gates,
  author       = {Zhenyu Li and Peng Liu and Peng Zhao and others},
  title        = {Error per single-qubit gate below 10\textsuperscript{-4} in a superconducting qubit},
  journal      = {npj Quantum Information},
  volume       = {9},
  pages        = {111},
  year         = {2023},
  doi          = {10.1038/s41534-023-00781-x},
  url          = {https://doi.org/10.1038/s41534-023-00781-x},
  publisher    = {Nature Publishing Group},
  received     = {2023-02-22},
  accepted     = {2023-10-22},
  published    = {2023-11-03}
}

@article{Jurcevic2021Gates,
  author       = {Petar Jurcevic and Ali Javadi-Abhari and Lev S. Bishop and Isaac Lauer and Daniela F. Bogorin and Markus Brink and Lauren Capelluto and Oktay G{\"u}nl{\"u}k and Toshinari Itoko and Naoki Kanazawa},
  title        = {Demonstration of quantum volume 64 on a superconducting quantum computing system},
  journal      = {Quantum Science and Technology},
  volume       = {6},
  number       = {2},
  pages        = {025020},
  year         = {2021},
  doi          = {10.1088/2058-9565/abe519},
  publisher    = {IOP Publishing},
  url          = {https://doi.org/10.1088/2058-9565/abe519}
}

@article{Sung2021Gates,
  title = {Realization of High-Fidelity CZ and $ZZ$-Free iSWAP Gates with a Tunable Coupler},
  author = {Sung, Youngkyu and Ding, Leon and Braum\"uller, Jochen and Veps\"al\"ainen, Antti and Kannan, Bharath and Kjaergaard, Morten and Greene, Ami and Samach, Gabriel O. and McNally, Chris and Kim, David and Melville, Alexander and Niedzielski, Bethany M. and Schwartz, Mollie E. and Yoder, Jonilyn L. and Orlando, Terry P. and Gustavsson, Simon and Oliver, William D.},
  journal = {Phys. Rev. X},
  volume = {11},
  issue = {2},
  pages = {021058},
  numpages = {32},
  year = {2021},
  month = {Jun},
  publisher = {American Physical Society},
  doi = {10.1103/PhysRevX.11.021058},
  url = {https://link.aps.org/doi/10.1103/PhysRevX.11.021058}
}

@article{Stehlik2021Gates,
  title = {Tunable Coupling Architecture for Fixed-Frequency Transmon Superconducting Qubits},
  author = {Stehlik, J. and Zajac, D. M. and Underwood, D. L. and Phung, T. and Blair, J. and Carnevale, S. and Klaus, D. and Keefe, G. A. and Carniol, A. and Kumph, M. and Steffen, Matthias and Dial, O. E.},
  journal = {Phys. Rev. Lett.},
  volume = {127},
  issue = {8},
  pages = {080505},
  numpages = {6},
  year = {2021},
  month = {Aug},
  publisher = {American Physical Society},
  doi = {10.1103/PhysRevLett.127.080505},
  url = {https://link.aps.org/doi/10.1103/PhysRevLett.127.080505}
}

@article{Wei2022Gates,
  title = {Hamiltonian Engineering with Multicolor Drives for Fast Entangling Gates and Quantum Crosstalk Cancellation},
  author = {Wei, K. X. and Magesan, E. and Lauer, I. and Srinivasan, S. and Bogorin, D. F. and Carnevale, S. and Keefe, G. A. and Kim, Y. and Klaus, D. and Landers, W. and Sundaresan, N. and Wang, C. and Zhang, E. J. and Steffen, M. and Dial, O. E. and McKay, D. C. and Kandala, A.},
  journal = {Phys. Rev. Lett.},
  volume = {129},
  issue = {6},
  pages = {060501},
  numpages = {6},
  year = {2022},
  month = {Aug},
  publisher = {American Physical Society},
  doi = {10.1103/PhysRevLett.129.060501},
  url = {https://link.aps.org/doi/10.1103/PhysRevLett.129.060501}
}

@article{Varbanov2020QEC,
  author    = {Varbanov, Borna M. and Battistel, Francesco and Tarasinski, Brian M. and Di Carlo, Leonardo and Terhal, Barbara M.},
  title     = {Leakage detection for a transmon-based surface code},
  journal   = {npj Quantum Information},
  volume    = {6},
  number    = {1},
  pages     = {102},
  year      = {2020},
  doi       = {10.1038/s41534-020-00330-w},
  url       = {https://doi.org/10.1038/s41534-020-00330-w},
  publisher = {Nature Publishing Group},
}

@article{Dumas2024,
  title = {Measurement-Induced Transmon Ionization},
  author = {Dumas, Marie Fr\'ed\'erique and Groleau-Par\'e, Benjamin and McDonald, Alexander and Mu\~noz-Arias, Manuel H. and Lled\'o, Crist\'obal and D'Anjou, Benjamin and Blais, Alexandre},
  journal = {Phys. Rev. X},
  volume = {14},
  issue = {4},
  pages = {041023},
  numpages = {31},
  year = {2024},
  month = {Oct},
  publisher = {American Physical Society},
  doi = {10.1103/PhysRevX.14.041023},
  url = {https://link.aps.org/doi/10.1103/PhysRevX.14.041023}
}

@misc{Wang2025exp,
  author       = {Zihao Wang and Benjamin D'Anjou and Philippe Gigon and Alexandre Blais and Machiel S. Blok},
  title        = {Probing excited-state dynamics of transmon ionization},
  year         = {2025},
  eprint       = {2505.00639},
  archivePrefix= {arXiv},
  primaryClass = {quant-ph},
  url          = {https://doi.org/10.48550/arXiv.2505.00639},
}

@misc{Fechant2025exp,
  author       = {Mathieu Féchant and Marie Frédérique Dumas and Denis Bénâtre and Nicolas Gosling and Philipp Lenhard and Martin Spiecker and Wolfgang Wernsdorfer and Benjamin D'Anjou and Alexandre Blais and Ioan M. Pop},
  title        = {Offset Charge Dependence of Measurement-Induced Transitions in Transmons},
  year         = {2025},
  eprint       = {2505.00674},
  archivePrefix= {arXiv},
  primaryClass = {quant-ph},
  url          = {https://doi.org/10.48550/arXiv.2505.00674},
}

@misc{Connolly2025multimode_exp,
  author       = {Thomas Connolly and Pavel D. Kurilovich and Vladislav D. Kurilovich and Charlotte G. L. Bøttcher and Sumeru Hazra and Wei Dai and Andy Z. Ding and Vidul R. Joshi and Heekun Nho and Spencer Diamond and Daniel K. Weiss and Valla Fatemi and Luigi Frunzio and Leonid I. Glazman and Michel H. Devoret},
  title        = {Full characterization of measurement-induced transitions of a superconducting qubit},
  year         = {2025},
  eprint       = {2506.05306},
  archivePrefix= {arXiv},
  primaryClass = {quant-ph},
  url          = {https://doi.org/10.48550/arXiv.2506.05306},
}

@ARTICLE{Xia2025-tj,
  title         = "Exceeding the parametric drive strength threshold in
                   nonlinear circuits",
  author        = "Xia, Mingkang and Lledó, Cristóbal and Capocci, Matthew and
                   Repicky, Jacob and D'Anjou, Benjamin and Mondragon-Shem, Ian
                   and Kaufman, Ryan and Koch, Jens and Blais, Alexandre and
                   Hatridge, Michael",
  journal       = "arXiv [quant-ph]",
  month         =  "3~" # jun,
  year          =  2025,
  archivePrefix = "arXiv",
  primaryClass  = "quant-ph"
}

@article{Nesterov_MIST_fluxonium,
  title = {Measurement-induced state transitions in dispersive qubit-readout schemes},
  author = {Nesterov, Konstantin N. and Pechenezhskiy, Ivan V.},
  journal = {Phys. Rev. Appl.},
  volume = {22},
  issue = {6},
  pages = {064038},
  numpages = {18},
  year = {2024},
  month = {Dec},
  publisher = {American Physical Society},
  doi = {10.1103/PhysRevApplied.22.064038},
  url = {https://link.aps.org/doi/10.1103/PhysRevApplied.22.064038}
}

@article{Chapple2025Balanced,
  title = {Balanced Cross-Kerr Coupling for Superconducting Qubit Readout},
  author = {Chapple, Alex A. and Benhayoune-Khadraoui, Othmane and Richer, Simon and Blais, Alexandre},
  journal = {Phys. Rev. Lett.},
  volume = {135},
  issue = {25},
  pages = {256002},
  numpages = {8},
  year = {2025},
  month = {Dec},
  publisher = {American Physical Society},
  doi = {10.1103/r4v5-wyyt},
  url = {https://link.aps.org/doi/10.1103/r4v5-wyyt}
}

@article{Chapple2025Robustness,
  title = {Robustness of longitudinal transmon readout to ionization},
  author = {Chapple, Alex A. and McDonald, Alexander and Mu\~noz-Arias, Manuel H. and Lachapelle, Mathieu and Blais, Alexandre},
  journal = {Phys. Rev. Appl.},
  volume = {24},
  issue = {3},
  pages = {034026},
  numpages = {16},
  year = {2025},
  month = {Sep},
  publisher = {American Physical Society},
  doi = {10.1103/mp2x-zj3y},
  url = {https://link.aps.org/doi/10.1103/mp2x-zj3y}
}

@ARTICLE{Manucharyan2009-Fluxonium,
  title    = "Fluxonium: single cooper-pair circuit free of charge offsets",
  author   = "Manucharyan, Vladimir E and Koch, Jens and Glazman, Leonid I and
              Devoret, Michel H",
  journal  = "Science",
  volume   =  326,
  number   =  5949,
  pages    = "113--116",
  month    =  "2~" # oct,
  year     =  2009,
  doi      = "10.1126/science.1175552",
}

@article{Brooks20130Pi,
  title = {Protected gates for superconducting qubits},
  author = {Brooks, Peter and Kitaev, Alexei and Preskill, John},
  journal = {Phys. Rev. A},
  volume = {87},
  issue = {5},
  pages = {052306},
  numpages = {26},
  year = {2013},
  month = {May},
  publisher = {American Physical Society},
  doi = {10.1103/PhysRevA.87.052306},
  url = {https://link.aps.org/doi/10.1103/PhysRevA.87.052306}
}

@ARTICLE{Chou2024-DualRail,
  title     = "A superconducting dual-rail cavity qubit with erasure-detected
               logical measurements",
  author    = "Chou, Kevin S and Shemma, Tali and McCarrick, Heather and Chien,
               Tzu-Chiao and Teoh, James D and Winkel, Patrick and Anderson,
               Amos and Chen, Jonathan and Curtis, Jacob C and de Graaf, Stijn J
               and Garmon, John W O and Gudlewski, Benjamin and Kalfus, William
               D and Keen, Trevor and Khedkar, Nishaad and Lei, Chan U and Liu,
               Gangqiang and Lu, Pinlei and Lu, Yao and Maiti, Aniket and
               Mastalli-Kelly, Luke and Mehta, Nitish and Mundhada, Shantanu O
               and Narla, Anirudh and Noh, Taewan and Tsunoda, Takahiro and Xue,
               Sophia H and Yuan, Joseph O and Frunzio, Luigi and Aumentado,
               José and Puri, Shruti and Girvin, Steven M and Moseley, Jr, S
               Harvey and Schoelkopf, Robert J",
  journal   = "Nat. Phys.",
  publisher = "Springer Science and Business Media LLC",
  volume    =  20,
  number    =  9,
  pages     = "1454--1460",
  month     =  "2~" # sep,
  year      =  2024,
  doi       = "10.1038/s41567-024-02539-4"
}

@article{Kubica2023ErasureQubits,
  title = {Erasure Qubits: Overcoming the T$_1$ Limit in Superconducting Circuits},
  author = {Kubica, Aleksander and Haim, Arbel and Vaknin, Yotam and Levine, Harry and Brandão, Fernando and Retzker, Alex},
  journal = {Physical Review X},
  volume = {13},
  number = {4},
  pages = {041022},
  year = {2023},
  doi = {10.1103/PhysRevX.13.041022}
}

@article{Dai2025DriveInduced,
  title   = {Characterization of drive-induced unwanted state transitions in superconducting circuits},
  author  = {Dai, Wei and Hazra, Sumeru and Weiss, Daniel K. and Kurilovich, Pavel D. and Connolly, Thomas and Babla, Harshvardhan K. and Singh, Shraddha and Joshi, Vidul R. and Ding, Andy Z. and Parakh, Pranav D. and Venkatraman, Jayameenakshi and Xiao, Xu and Frunzio, Luigi and Devoret, Michel H.},
  journal = {Physical Review X},
  year    = {2025},
  note    = {Published online},
  doi     = {10.1103/PhysRevX.16.011011}
}

@article{Yan2018TunableCoupler,
  title = {Tunable Coupling Scheme for Implementing High-Fidelity Two-Qubit Gates},
  author = {Yan, Fei and Krantz, Philip and Sung, Youngkyu and Kjaergaard, Morten and Campbell, Daniel L. and Orlando, Terry P. and Gustavsson, Simon and Oliver, William D.},
  journal = {Phys. Rev. Appl.},
  volume = {10},
  issue = {5},
  pages = {054062},
  numpages = {9},
  year = {2018},
  month = {Nov},
  publisher = {American Physical Society},
  doi = {10.1103/PhysRevApplied.10.054062},
  url = {https://link.aps.org/doi/10.1103/PhysRevApplied.10.054062}
}

@article{Hone1997gap,
  title = {Time-dependent Floquet theory and absence of an adiabatic limit},
  author = {Hone, Daniel W. and Ketzmerick, Roland and Kohn, Walter},
  journal = {Phys. Rev. A},
  volume = {56},
  issue = {5},
  pages = {4045--4054},
  numpages = {0},
  year = {1997},
  month = {Nov},
  publisher = {American Physical Society},
  doi = {10.1103/PhysRevA.56.4045},
  url = {https://link.aps.org/doi/10.1103/PhysRevA.56.4045}
}

@article{Leghtas2013Cat,
  title = {Hardware-Efficient Autonomous Quantum Memory Protection},
  author = {Leghtas, Zaki and Kirchmair, Gerhard and Vlastakis, Brian and Schoelkopf, Robert J. and Devoret, Michel H. and Mirrahimi, Mazyar},
  journal = {Phys. Rev. Lett.},
  volume = {111},
  issue = {12},
  pages = {120501},
  numpages = {5},
  year = {2013},
  month = {Sep},
  publisher = {American Physical Society},
  doi = {10.1103/PhysRevLett.111.120501},
  url = {https://link.aps.org/doi/10.1103/PhysRevLett.111.120501}
}

@article{Grimm2020Cat,
  author        = {Grimm, A. and Frattini, N. E. and Puri, S. and Mundhada, S. O. and Touzard, S. and Mirrahimi, M. and Girvin, S. M. and Shankar, S. and Devoret, M. H.},
  date          = {2020/08/01},
  doi           = {10.1038/s41586-020-2587-z},
  id            = {Grimm2020},
  isbn          = {1476-4687},
  journal       = {Nature},
  number        = {7820},
  pages         = {205--209},
  title         = {Stabilization and operation of a Kerr-cat qubit},
  url           = {https://doi.org/10.1038/s41586-020-2587-z},
  volume        = {584},
  year          = {2020}
}

@article{CampagneIbarcq2020GKP,
  date          = {2020/08/01},
  doi           = {10.1038/s41586-020-2603-3},
  id            = {Campagne-Ibarcq2020},
  isbn          = {1476-4687},
  journal       = {Nature},
  number        = {7821},
  pages         = {368--372},
  title         = {Quantum error correction of a qubit encoded in grid states of an oscillator},
  url           = {https://doi.org/10.1038/s41586-020-2603-3},
  volume        = {584},
  year          = {2020}
}

@article{Sivak2023GKP,
  date          = {2023/04/01},
  doi           = {10.1038/s41586-023-05782-6},
  id            = {Sivak2023},
  isbn          = {1476-4687},
  journal       = {Nature},
  number        = {7955},
  pages         = {50--55},
  title         = {Real-time quantum error correction beyond break-even},
  url           = {https://doi.org/10.1038/s41586-023-05782-6},
  volume        = {616},
  year          = {2023}
}

@article{Schreier2008Transmon,
  title = {Suppressing charge noise decoherence in superconducting charge qubits},
  author = {Schreier, J. A. and Houck, A. A. and Koch, Jens and Schuster, D. I. and Johnson, B. R. and Chow, J. M. and Gambetta, J. M. and Majer, J. and Frunzio, L. and Devoret, M. H. and Girvin, S. M. and Schoelkopf, R. J.},
  journal = {Phys. Rev. B},
  volume = {77},
  issue = {18},
  pages = {180502},
  numpages = {4},
  year = {2008},
  month = {May},
  publisher = {American Physical Society},
  doi = {10.1103/PhysRevB.77.180502},
  url = {https://link.aps.org/doi/10.1103/PhysRevB.77.180502}
}

@article{Nguyen2019Fluxonium,
  title     = {High-Coherence Fluxonium Qubit},
  author    = {Nguyen, Long B. and Lin, Yen-Hsiang and Somoroff, Aaron and Mencia, Raymond and Grabon, Nicholas and Manucharyan, Vladimir E.},
  journal   = prx,
  volume    = {9},
  issue     = {4},
  pages     = {041041},
  numpages  = {14},
  year      = {2019},
  month     = {11},
  publisher = {American Physical Society},
  url       = {https://link.aps.org/doi/10.1103/PhysRevX.9.041041}
}

@article{Barends2013Xmon,
  author     = {Barends, R. and Kelly, J. and Megrant, A. and Sank, D. and Jeffrey, E. and Chen, Y. and Yin, Y. and Chiaro, B. and Mutus, J. and Neill, C. and O'Malley, P. and Roushan, P. and Wenner, J. and White, T. C. and Cleland, A. N. and Martinis, John M.},
  doi        = {10.1103/PhysRevLett.111.080502},
  issue      = {8},
  journal    = {Phys. Rev. Lett.},
  month      = {8},
  numpages   = {5},
  pages      = {080502},
  publisher  = {American Physical Society},
  title      = {Coherent {J}osephson Qubit Suitable for Scalable Quantum Integrated Circuits},
  url        = {http://link.aps.org/doi/10.1103/PhysRevLett.111.080502},
  volume     = {111},
  year       = {2013}
}

@article{Paik2011Transmon3D,
  author    = {Paik, Hanhee and Schuster, D. I. and Bishop, Lev S. and Kirchmair, G. and Catelani, G. and Sears, A. P. and Johnson, B. R. and Reagor, M. J. and Frunzio, L. and Glazman, L. I. and Girvin, S. M. and Devoret, M. H. and Schoelkopf, R. J.},
  issue     = {24},
  journal   = prl,
  numpages  = {5},
  pages     = {240501},
  publisher = {American Physical Society},
  title     = {Observation of high coherence in {J}osephson junction qubits measured in a three-dimensional circuit {QED} architecture},
  volume    = {107},
  year      = {2011}
}

@article{Gyenis20210pi,
  title     = {Experimental Realization of a Protected Superconducting Circuit Derived from the $0$--$\ensuremath{\pi}$ Qubit},
  author    = {Gyenis, Andr\'as and Mundada, Pranav S. and Di Paolo, Agustin and Hazard, Thomas M. and You, Xinyuan and Schuster, David I. and Koch, Jens and Blais, Alexandre and Houck, Andrew A.},
  journal   = {PRX Quantum},
  volume    = {2},
  issue     = {1},
  pages     = {010339},
  numpages  = {18},
  year      = {2021},
  month     = {3},
  publisher = {American Physical Society},
  url       = {https://link.aps.org/doi/10.1103/PRXQuantum.2.010339}
}

@article{Barends2014Gates,
  author   = {Barends, R and Kelly, J and Megrant, A and Veitia, A and Sank, D and Jeffrey, E and White, T C and Mutus, J and Fowler, A G and Campbell, B and Chen, Y and Chen, Z and Chiaro, B and Dunsworth, A and Neill, C and O'Malley, P and Roushan, P and Vainsencher, A and Wenner, J and Korotkov, A N and Cleland, A N and Martinis, John M},
  issn     = {1476-4687},
  journal  = nat,
  month    = apr,
  number   = {7497},
  pages    = {500},
  title    = {{Superconducting quantum circuits at the surface code threshold for fault tolerance.}},
  volume   = {508},
  year     = {2014},
  url      = {http://www.nature.com/nature/journal/v508/n7497/abs/nature13171.html}
}

@article{Rol16SQGate,
  title     = {Restless Tuneup of High-Fidelity Qubit Gates},
  author    = {Rol, M. A. and Bultink, C. C. and O'Brien, T. E. and de Jong, S. R. and Theis, L. S. and Fu, X. and Luthi, F. and Vermeulen, R. F. L. and de Sterke, J. C. and Bruno, A. and Deurloo, D. and Schouten, R. N. and Wilhelm, F. K. and DiCarlo, L.},
  journal   = {Phys. Rev. Applied},
  volume    = {7},
  issue     = {4},
  pages     = {041001},
  numpages  = {6},
  year      = {2017},
  month     = {Apr},
  publisher = {American Physical Society},
  doi       = {10.1103/PhysRevApplied.7.041001},
  url       = {https://link.aps.org/doi/10.1103/PhysRevApplied.7.041001}
}

@article{Barends2019Diabatic,
  title     = {Diabatic Gates for Frequency-Tunable Superconducting Qubits},
  author    = {Barends, R. and Quintana, C. M. and Petukhov, A. G. and Chen, Yu and Kafri, D. and Kechedzhi, K. and Collins, R. and Naaman, O. and Boixo, S. and Arute, F. and Arya, K. and Buell, D. and Burkett, B. and Chen, Z. and Chiaro, B. and Dunsworth, A. and Foxen, B. and Fowler, A. and Gidney, C. and Giustina, M. and Graff, R. and Huang, T. and Jeffrey, E. and Kelly, J. and Klimov, P. V. and Kostritsa, F. and Landhuis, D. and Lucero, E. and McEwen, M. and Megrant, A. and Mi, X. and Mutus, J. and Neeley, M. and Neill, C. and Ostby, E. and Roushan, P. and Sank, D. and Satzinger, K. J. and Vainsencher, A. and White, T. and Yao, J. and Yeh, P. and Zalcman, A. and Neven, H. and Smelyanskiy, V. N. and Martinis, John M.},
  journal   = prl,
  volume    = {123},
  issue     = {21},
  pages     = {210501},
  numpages  = {6},
  year      = {2019},
  month     = {Nov},
  publisher = {American Physical Society},
  url       = {https://link.aps.org/doi/10.1103/PhysRevLett.123.210501}
}

@article{Rol2019Netzero,
  author  = {M. A. Rol and F. Battistel and F. K. Malinowski and C. C. Bultink and B. M. Tarasinski and R. Vollmer and N. Haider and N. Muthusubramanian and A. Bruno and B. M. Terhal and L. DiCarlo},
  title   = {Fast, High-Fidelity Conditional-Phase Gate Exploiting Leakage Interference in Weakly Anharmonic Superconducting Qubits},
  journal = prl,
  volume  = {123},
  pages   = {120502},
  year    = {2019},
  url     = {https://journals.aps.org/prl/abstract/10.1103/PhysRevLett.123.120502},
  arxivid = {1903.02492}
}

@article{Negirneac2021FastNetZero,
  title     = {High-Fidelity Controlled-$Z$ Gate with Maximal Intermediate Leakage Operating at the Speed Limit in a Superconducting Quantum Processor},
  author    = {Neg\^{\i}rneac, V. and Ali, H. and Muthusubramanian, N. and Battistel, F. and Sagastizabal, R. and Moreira, M. S. and Marques, J. F. and Vlothuizen, W. J. and Beekman, M. and Zachariadis, C. and Haider, N. and Bruno, A. and DiCarlo, L.},
  journal   = {Phys. Rev. Lett.},
  volume    = {126},
  issue     = {22},
  pages     = {220502},
  numpages  = {6},
  year      = {2021},
  month     = {Jun},
  publisher = {American Physical Society},
  doi       = {10.1103/PhysRevLett.126.220502},
  url       = {https://link.aps.org/doi/10.1103/PhysRevLett.126.220502}
}

@article{Foxen2020Continious,
  title         = {Demonstrating a Continuous Set of Two-Qubit Gates for Near-Term Quantum Algorithms},
  author        = {Foxen, B. and Neill, C. and Dunsworth, A. and Roushan, P. and Chiaro, B. and Megrant, A. and Kelly, J. and Chen, Zijun and Satzinger, K. and Barends, R. and Arute, F. and Arya, K. and Babbush, R. and Bacon, D. and Bardin, J. C. and Boixo, S. and Buell, D. and Burkett, B. and Chen, Yu and Collins, R. and Farhi, E. and Fowler, A. and Gidney, C. and Giustina, M. and Graff, R. and Harrigan, M. and Huang, T. and Isakov, S. V. and Jeffrey, E. and Jiang, Z. and Kafri, D. and Kechedzhi, K. and Klimov, P. and Korotkov, A. and Kostritsa, F. and Landhuis, D. and Lucero, E. and McClean, J. and McEwen, M. and Mi, X. and Mohseni, M. and Mutus, J. Y. and Naaman, O. and Neeley, M. and Niu, M. and Petukhov, A. and Quintana, C. and Rubin, N. and Sank, D. and Smelyanskiy, V. and Vainsencher, A. and White, T. C. and Yao, Z. and Yeh, P. and Zalcman, A. and Neven, H. and Martinis, J. M.},
  collaboration = {Google AI Quantum},
  journal       = {Phys. Rev. Lett.},
  volume        = {125},
  issue         = {12},
  pages         = {120504},
  numpages      = {6},
  year          = {2020},
  month         = {Sep},
  publisher     = {American Physical Society},
  doi           = {10.1103/PhysRevLett.125.120504},
  url           = {https://link.aps.org/doi/10.1103/PhysRevLett.125.120504}
}

@article{Hong2019Parametric,
  title     = {Demonstration of a parametrically activated entangling gate protected from flux noise},
  author    = {Hong, Sabrina S. and Papageorge, Alexander T. and Sivarajah, Prasahnt and Crossman, Genya and Didier, Nicolas and Polloreno, Anthony M. and Sete, Eyob A. and Turkowski, Stefan W. and da Silva, Marcus P. and Johnson, Blake R.},
  journal   = pra,
  volume    = {101},
  issue     = {1},
  pages     = {012302},
  numpages  = {8},
  year      = {2020},
  month     = {Jan},
  publisher = {American Physical Society},
  url       = {https://link.aps.org/doi/10.1103/PhysRevA.101.012302}
}

@article{Marshall2025Incoherent,
  title = {Incoherent approximation of leakage in quantum error correction},
  author = {Marshall, Jeffrey and Kafri, Dvir},
  journal = {Phys. Rev. Appl.},
  volume = {23},
  issue = {5},
  pages = {054025},
  numpages = {24},
  year = {2025},
  month = {May},
  publisher = {American Physical Society},
  doi = {10.1103/PhysRevApplied.23.054025},
  url = {https://link.aps.org/doi/10.1103/PhysRevApplied.23.054025}
}

@article{McEwen2021,
  author        = {McEwen, M. and Kafri, D. and Chen, Z. and Atalaya, J. and Satzinger, K. J. and Quintana, C. and Klimov, P. V. and Sank, D. and Gidney, C. and Fowler, A. G. and Arute, F. and Arya, K. and Buckley, B. and Burkett, B. and Bushnell, N. and Chiaro, B. and Collins, R. and Demura, S. and Dunsworth, A. and Erickson, C. and Foxen, B. and Giustina, M. and Huang, T. and Hong, S. and Jeffrey, E. and Kim, S. and Kechedzhi, K. and Kostritsa, F. and Laptev, P. and Megrant, A. and Mi, X. and Mutus, J. and Naaman, O. and Neeley, M. and Neill, C. and Niu, M. and Paler, A. and Redd, N. and Roushan, P. and White, T. C. and Yao, J. and Yeh, P. and Zalcman, A. and Chen, Yu and Smelyanskiy, V. N. and Martinis, John M. and Neven, H. and Kelly, J. and Korotkov, A. N. and Petukhov, A. G. and Barends, R.},
  date          = {2021/03/19},
  doi           = {10.1038/s41467-021-21982-y},
  id            = {McEwen2021},
  isbn          = {2041-1723},
  journal       = {Nature Communications},
  number        = {1},
  pages         = {1761},
  title         = {Removing leakage-induced correlated errors in superconducting quantum error correction},
  url           = {https://doi.org/10.1038/s41467-021-21982-y},
  volume        = {12},
  year          = {2021}
}

@article{Miao2022,
  author        = {Miao, Kevin C. and McEwen, Matt and Atalaya, Juan and Kafri, Dvir and Pryadko, Leonid P. and Bengtsson, Andreas and Opremcak, Alex and Satzinger, Kevin J. and Chen, Zijun and Klimov, Paul V. and Quintana, Chris and Acharya, Rajeev and Anderson, Kyle and Ansmann, Markus and Arute, Frank and Arya, Kunal and Asfaw, Abraham and Bardin, Joseph C. and Bourassa, Alexandre and Bovaird, Jenna and Brill, Leon and Buckley, Bob B. and Buell, David A. and Burger, Tim and Burkett, Brian and Bushnell, Nicholas and Campero, Juan and Chiaro, Ben and Collins, Roberto and Conner, Paul and Crook, Alexander L. and Curtin, Ben and Debroy, Dripto M. and Demura, Sean and Dunsworth, Andrew and Erickson, Catherine and Fatemi, Reza and Ferreira, Vinicius S. and Burgos, Leslie Flores and Forati, Ebrahim and Fowler, Austin G. and Foxen, Brooks and Garcia, Gonzalo and Giang, William and Gidney, Craig and Giustina, Marissa and Gosula, Raja and Dau, Alejandro Grajales and Gross, Jonathan A. and Hamilton, Michael C. and Harrington, Sean D. and Heu, Paula and Hilton, Jeremy and Hoffmann, Markus R. and Hong, Sabrina and Huang, Trent and Huff, Ashley and Iveland, Justin and Jeffrey, Evan and Jiang, Zhang and Jones, Cody and Kelly, Julian and Kim, Seon and Kostritsa, Fedor and Kreikebaum, John Mark and Landhuis, David and Laptev, Pavel and Laws, Lily and Lee, Kenny and Lester, Brian J. and Lill, Alexander T. and Liu, Wayne and Locharla, Aditya and Lucero, Erik and Martin, Steven and Megrant, Anthony and Mi, Xiao and Montazeri, Shirin and Morvan, Alexis and Naaman, Ofer and Neeley, Matthew and Neill, Charles and Nersisyan, Ani and Newman, Michael and Ng, Jiun How and Nguyen, Anthony and Nguyen, Murray and Potter, Rebecca and Rocque, Charles and Roushan, Pedram and Sankaragomathi, Kannan and Schurkus, Henry F. and Schuster, Christopher and Shearn, Michael J. and Shorter, Aaron and Shutty, Noah and Shvarts, Vladimir and Skruzny, Jindra and Smith, W. Clarke and Sterling, George and Szalay, Marco and Thor, Douglas and Torres, Alfredo and White, Theodore and Woo, Bryan W. K. and Yao, Z. Jamie and Yeh, Ping and Yoo, Juhwan and Young, Grayson and Zalcman, Adam and Zhu, Ningfeng and Zobrist, Nicholas and Neven, Hartmut and Smelyanskiy, Vadim and Petukhov, Andre and Korotkov, Alexander N. and Sank, Daniel and Chen, Yu},
  date          = {2023/10/05},
  doi           = {10.1038/s41567-023-02226-w},
  id            = {Miao2023},
  isbn          = {1745-2481},
  journal       = {Nature Physics},
  title         = {Overcoming leakage in quantum error correction},
  url           = {https://doi.org/10.1038/s41567-023-02226-w},
  year          = {2023}
}

@article{Nesterov2024MIST,
  title = {Measurement-induced state transitions in dispersive qubit-readout schemes},
  author = {Nesterov, Konstantin N. and Pechenezhskiy, Ivan V.},
  journal = {Phys. Rev. Appl.},
  volume = {22},
  issue = {6},
  pages = {064038},
  numpages = {18},
  year = {2024},
  month = {Dec},
  publisher = {American Physical Society},
  doi = {10.1103/PhysRevApplied.22.064038},
  url = {https://link.aps.org/doi/10.1103/PhysRevApplied.22.064038}
}

@article{Aliferis07,
  author     = {Aliferis, Panos and Terhal, Barbara M.},
  title      = {Fault-tolerant Quantum Computation for Local Leakage Faults},
  journal    = {Quantum Info. Comput.},
  issue_date = {January 2007},
  volume     = {7},
  number     = {1},
  month      = jan,
  year       = {2007},
  issn       = {1533-7146},
  pages      = {139--156},
  numpages   = {18},
  url        = {http://dl.acm.org/citation.cfm?id=2011706.2011715},
  acmid      = {2011715},
  publisher  = {Rinton Press, Incorporated},
  address    = {Paramus, NJ}
}

@article{Fowler13,
  title    = {Coping with qubit leakage in topological codes},
  author   = {Fowler, Austin G.},
  journal  = {Phys. Rev. A},
  volume   = {88},
  issue    = {4},
  pages    = {042308},
  numpages = {5},
  year     = {2013},
  url      = {https://link.aps.org/doi/10.1103/PhysRevA.88.042308}
}

@article{Ghosh13b,
  title     = {Understanding the effects of leakage in superconducting quantum-error-detection circuits},
  author    = {Ghosh, Joydip and Fowler, Austin G. and Martinis, John M. and Geller, Michael R.},
  journal   = {Phys. Rev. A},
  volume    = {88},
  issue     = {6},
  pages     = {062329},
  numpages  = {7},
  year      = {2013},
  month     = {12},
  publisher = {American Physical Society},
  doi       = {10.1103/PhysRevA.88.062329},
  url       = {https://link.aps.org/doi/10.1103/PhysRevA.88.062329}
}

@article{Ghosh15,
  title     = {Leakage-resilient approach to fault-tolerant quantum computing with superconducting elements},
  author    = {Ghosh, Joydip and Fowler, Austin G.},
  journal   = {Phys. Rev. A},
  volume    = {91},
  issue     = {2},
  pages     = {020302(R)},
  numpages  = {5},
  year      = {2015},
  month     = {2},
  publisher = {American Physical Society},
  doi       = {10.1103/PhysRevA.91.020302},
  url       = {https://link.aps.org/doi/10.1103/PhysRevA.91.020302}
}

@article{Kelly15,
  author    = {Kelly, J and Barends, R and Fowler, A. G. and Megrant, A and Jeffrey, E and White, TC and Sank, D and Mutus, JY and Campbell, B and Chen, Yu and Chiaro, B. and Dunsworth , A. and Hoi , I.-C. and Neill , C. and O’Malley , P. J. J. and Quintana , C. and Roushan , P. and Vainsencher , A. and Cleland , J. Wenner, A. N. and Martinis, John M.},
  title     = {State preservation by repetitive error detection in a superconducting quantum circuit},
  number    = {7541},
  pages     = {66--69},
  url       = {https://www.nature.com/nature/journal/v519/n7541/full/nature14270.html},
  volume    = {519},
  journal   = nat,
  publisher = {Nature Publishing Group},
  year      = {2015}
}

@article{Suchara15,
  author     = {Suchara, Martin and Cross, Andrew W. and Gambetta, Jay M.},
  title      = {Leakage Suppression in the Toric Code},
  journal    = {Quantum Info. Comput.},
  issue_date = {September 2015},
  volume     = {15},
  number     = {11-12},
  month      = sep,
  year       = {2015},
  issn       = {1533-7146},
  pages      = {997--1016},
  numpages   = {20},
  url        = {http://dl.acm.org/citation.cfm?id=2871350.2871358},
  acmid      = {2871358},
  publisher  = {Rinton Press, Incorporated},
  address    = {Paramus, NJ}
}

@article{Boissonneault2010_ImproveNonlinear,
  title = {Improved Superconducting Qubit Readout by Qubit-Induced Nonlinearities},
  author = {Boissonneault, Maxime and Gambetta, J. M. and Blais, Alexandre},
  journal = {Phys. Rev. Lett.},
  volume = {105},
  issue = {10},
  pages = {100504},
  numpages = {4},
  year = {2010},
  month = {Sep},
  publisher = {American Physical Society},
  doi = {10.1103/PhysRevLett.105.100504},
  url = {https://link.aps.org/doi/10.1103/PhysRevLett.105.100504}
}

@misc{xiao2024diagrammaticmethodcomputeeffective,
      title={A diagrammatic method to compute the effective Hamiltonian of driven nonlinear oscillators}, 
      author={Xu Xiao and Jayameenakshi Venkatraman and Rodrigo G. Cortiñas and Shoumik Chowdhury and Michel H. Devoret},
      year={2024},
      eprint={2304.13656},
      archivePrefix={arXiv},
      primaryClass={quant-ph},
      url={https://arxiv.org/abs/2304.13656}, 
}

@misc{Chapple2026FluxoniumMIST,
  author       = {Chapple, Alex A. and Varbanov, Boris M. and McDonald, Alexander and Blais, Alexandre},
  title        = {Measurement-induced state transitions across the fluxonium qubit landscape},
  year         = {2026},
  eprint       = {2604.08515},
  archivePrefix= {arXiv},
  primaryClass = {quant-ph},
  note         = {arXiv:2604.08515v1}
}

@article{Heinsoo2018readout,
  title = {Rapid High-fidelity Multiplexed Readout of Superconducting Qubits},
  author = {Heinsoo, Johannes and Andersen, Christian Kraglund and Remm, Ants and Krinner, Sebastian and Walter, Theodore and Salath\'e, Yves and Gasparinetti, Simone and Besse, Jean-Claude and Poto\ifmmode \check{c}\else \v{c}\fi{}nik, Anton and Wallraff, Andreas and Eichler, Christopher},
  journal = {Phys. Rev. Appl.},
  volume = {10},
  issue = {3},
  pages = {034040},
  numpages = {14},
  year = {2018},
  month = {Sep},
  publisher = {American Physical Society},
  doi = {10.1103/PhysRevApplied.10.034040},
  url = {https://link.aps.org/doi/10.1103/PhysRevApplied.10.034040}
}

\end{document}